\documentclass{aa}  
\usepackage{color}
\usepackage{subcaption}
\usepackage{multirow}

\usepackage{graphicx}
\usepackage{txfonts}
\begin{document}

   \title{B fields in OB stars (BOB):}

   \subtitle{Detection of a strong magnetic field in the O9.7~V star HD 54879\thanks{Based on observations made with ESO telescopes at the La Silla and Paranal observatories under programme ID 191.D-0255(C,F).}}

   \author{N.~Castro\inst{1},
                   L.~Fossati\inst{1},
                   S.~Hubrig\inst{2},
                   S.~Simón-Díaz\inst{3,4},                                   
                   M.~Sch\"oller\inst{5},
                   I.~Ilyin\inst{2},
                   T.~A.~Carrol\inst{2},
                   N.~Langer\inst{1},
                   T.~Morel\inst{6},               
                   F.~R.~N.~Schneider\inst{1,7},
                   N.~Przybilla\inst{8},
                   A.~Herrero\inst{3,4},
                   A.~de~Koter\inst{9,10},                         
                   L.~M.~Oskinova\inst{11}, 
                   A.~Reisenegger\inst{12}, 
                   H.~Sana\inst{13} 
                   and the BOB collaboration       
          }
     \institute{Argelander-Institut für Astronomie der Universität Bonn, Auf dem Hügel 71, 53121, Bonn, Germany\\
              \email{norberto@astro.uni-bonn.de}
         \and
         Leibniz-Institut für Astrophysik Potsdam (AIP), An der Sternwarte 16, D-14482 Potsdam, Germany
         \and         
            Instituto de Astrofísica de Canarias, 38200, La Laguna, Tenerife, Spain 
       \and       
        Universidad de La Laguna, 38205, La Laguna, Tenerife, Spain
                \and        
         European Southern Observatory, Karl-Schwarzschild-Str.~2, 85748 Garching bei M\"unchen, Germany 
                \and          
        Institut d'Astrophysique et de Géophysique, Université de Liège, Allée du 6 Août, Bât. B5c, 4000, Liège, Belgium
                \and
    Department of Physics, University of Oxford, Denys Wilkinson Building, Keble Road, Oxford OX1 3RH, United Kingdom          
       \and
                Institut für Astro- und Teilchenphysik, Universität Innsbruck, Technikerstr. 25/8, 6020, Innsbruck, Austria                   
                \and
                Astronomical Institute Anton Pannekoek, University of Amsterdam, Kruislaan 403, 1098 SJ, Amsterdam, The Netherlands
                \and            
                Instituut voor Sterrenkunde, KU Leuven, Celestijnenlaan 200D, 3001 Leuven, Belgium    
                \and
                Institute for Physics and Astronomy, University of Potsdam, D-14476 Potsdam, Germany 
                \and
                Instituto de Astrofísica, Pontificia Universidad Católica de Chile, Casilla 306, Santiago 22, Chile  
                \and             
        European Space Agency, Space Telescope Science Institute, 3700 San Martin Drive, Baltimore, MD 21218, USA      
             }

   \date{Received --; accepted --}
\titlerunning{First magnetic detection in the O9.7~V star HD\,54879}
\authorrunning{N. Castro et al.}
 
  \abstract
   {The  number of magnetic stars detected among massive stars is small; nevertheless, the role played by the magnetic field in stellar evolution cannot be disregarded. Links between line profile variability, enhancements/depletions of surface chemical abundances, and magnetic fields have been identified for low-mass B-stars, but for the O-type domain this is almost unexplored. Based on  FORS\,2 and HARPS spectropolarimetric data, we present the first detection of a magnetic field in HD\,54879, a single slowly rotating O9.7~V star. Using two independent and different techniques we obtained the firm detection of a surface average longitudinal magnetic field with a maximum amplitude of about 600\,G, in modulus. A quantitative spectroscopic analysis of the star with the stellar atmosphere code {\sc fastwind} results in an effective temperature and a surface gravity of 33000$\pm1000$\,K and 4.0$\pm0.1$\,dex. The abundances of carbon, nitrogen, oxygen, silicon, and magnesium are found to be slightly lower than solar, but compatible within the errors. We investigate line-profile variability in HD\,54879 by complementing our spectra with  spectroscopic data from other recent OB-star surveys. The photospheric lines remain constant in shape between 2009 and 2014, although H$\alpha$ shows a variable emission. The H$\alpha$ emission is too strong for a standard O9.7~V and is probably linked to the magnetic field and the presence of circumstellar material. Its normal chemical composition and the absence of photospheric line profile variations make HD\,54879  the most strongly magnetic, non-variable single O-star detected to date. }
 
 
   \maketitle

\section{Introduction}

The nature and role of magnetic fields in massive stars is currently poorly understood \citep{2000ARA&A..38..143M,2009MNRAS.392.1022U,2012MNRAS.427..483B,2012ARA&A..50..107L}.  The origin of magnetic fields in main-sequence stars  in the upper part of the Hertzsprung-Russell diagram (HRD) is being debated \citep{2001ASPC..248..305M,2009MNRAS.400L..71F,2010ARep...54..156T,2014IAUS..302....1L,2014MNRAS.437..675W} and the role that these fields play in stellar evolution is being explored \citep{2012ARA&A..50..107L}.

The properties of magnetic fields in  intermediate mass stars (A and late-B stars) have been studied in detail \citep[see][and references therein]{1992A&ARv...4...35L,2006A&A...450..763K,
2007AN....328..475H,2009ARA&A..47..333D},  presenting links between the magnetic field strength, chemical peculiarities, and age \citep[Ap/Bp stars,][]{1980ApJS...42..421B,2007A&A...470..685L,2008A&A...481..465L,2014A&A...561A.147B}. However, magnetism in the upper part of the HRD is  poorly understood \citep{2009ARA&A..47..333D,2012ARA&A..50..107L,2012A&A...538A..29M}.  Magnetic fields have been unambiguously detected only in a few dozen main-sequence O- and B-type stars \citep[see e.g.][and references therein]{2013A&A...557L..16B,2013MNRAS.429..398P,2014arXiv1404.5508A,2014A&A...562A.143F,     Fossati2014,2014A&A...564L..10H,2014A&A...563L...7N}, revealing a small magnetic field incidence of 7\% \citep{2013arXiv1310.3965W,2015IAUS..307..342M}. In the O-star subsample, the confirmed magnetic field detections are even scarcer; only ten stars have  been reported \citep{2002MNRAS.333...55D,
2006MNRAS.365L...6D,
2008MNRAS.389...75B,
2008A&A...490..793H,
2009MNRAS.400L..94G,
2010MNRAS.407.1423M,
2011A&A...528A.151H,
2012MNRAS.423.3413N,
2012MNRAS.419.2459W,
2012MNRAS.425.1278W,
2013MNRAS.428.1686G}, most of them classified as Of?p stars or peculiar\footnote{We consider peculiar stars those objects that show any uncommon spectral feature compared to the standard spectral type classification criteria \citep{1990PASP..102..379W,2000PASP..112...50W,2011ApJS..193...24S}, for instance, the presence of C\,{\sc iii} $\lambda\lambda4647-4650-4652\,\AA$ emission lines \citep{1972AJ.....77..312W}. We do not consider H$\alpha$ as part of these classification criteria.} in some other way (e.g. $\theta^1$ Ori~C is  a well-known variable star).

A larger number of detections is mandatory to find links between  stellar parameters and magnetic fields. The ``B fields in OB stars'' (BOB) collaboration methodically searches for magnetic fields in slowly rotating, massive main-sequence stars \citep{2014arXiv1408.2100M,2015IAUS..307..342M}. The first results from the BOB project have been published by \cite{2014A&A...564L..10H,2015A&A...578L...3H}  and \cite{Fossati2014}.

We present here the first detection of a  strong magnetic field in the star HD\,54879 and the quantitative characterisation of its optical spectrum. HD\,54879 \citep[$V=7.65$,][]{2003AJ....125.2531R} is an O9.7~V star \citep{2011ApJS..193...24S} and a probable member of the CMa OB1 association, which is about 3\,Myr old  \citep{1974A&A....37..229C}, at  a distance of $\sim1320\pm130$\,pc  \citep{1978ApJS...38..309H}. In the recent catalogue of projected rotational velocities of northern O- and early B-type stars \citep{2014A&A...562A.135S}, this star is quoted as having a low projected rotational velocity (\ensuremath{{\upsilon}\sin i}) and macroturbulent broadening component of 6 and 10 km\,s$^{-1}$, respectively.

The paper is organised as follows. Section 2 describes the observational material. Section 3 presents the results of the magnetic field detection and the analysis techniques. Section 4 details the  analysis of the stellar atmosphere and the procedures employed. In Sect. 5 we discuss our results and  the conclusions are drawn in Sect. 6.

\section{Observations}  
\label{Obs}
We observed HD\,54879 using the FORS\,2 low-resolution spectropolarimeter \citep{1992Msngr..67...18A,1998Msngr..94....1A} attached to the Cassegrain focus of the 8\,m Antu telescope of the ESO Very Large Telescope of the Paranal Observatory. The observations were performed using the 2k$\times$4k MIT CCDs (pixel size 15\,$\mu$m\,$\times$\,15\,$\mu$m) and a narrow slit with a width of 0.4$\arcsec$, leading to a (measured) resolving power of about 1700. We also adopted the 200\,kHz/low/1$\times$1 readout mode and the GRISM\,600B. Each spectrum covers the $3250-6215$\,\AA\ spectral range which includes all Balmer lines, except H$\alpha$, and a number of He lines. The star was observed on two consecutive nights (Feb. 7 and 8, 2014) with a sequence of spectra obtained by  rotating the quarter waveplate from $-$45$^{\circ}$ to $+$45$^{\circ}$ every second exposure (i.e., $-$45$^{\circ}$, $+$45$^{\circ}$, $+$45$^{\circ}$, $-$45$^{\circ}$, $-$45$^{\circ}$, $+$45$^{\circ}$, etc.). The exposure times and signal-to-noise ratio (S/N) of Stokes $I$ are listed in Table~\ref{tab:fors}.

We also observed HD\,54879 using the HARPSpol polarimeter \citep{2011ASPC..437..237S,2011Msngr.143....7P} feeding the HARPS spectrograph \citep{2003Msngr.114...20M} attached to the ESO 3.6\,m telescope in La\,Silla, Chile. The observations, covering the $3780-6910$\,\AA\ wavelength range with a resolving power $R$\,$\sim$115000, were obtained on April 23, 2014, using the circular polarisation analyser. We observed the star with a sequence of four subexposures obtained rotating the quarter-wave retarder plate by 90$^\circ$ after each exposure, i.e. 45$^\circ$, 135$^\circ$, 225$^\circ$, and 315$^\circ$. Each subexposure was acquired using an exposure time of 2700\,seconds, leading to a Stokes $I$ S/N per pixel of 347 at  $\sim$\,$4950\,\AA$.

\section{Magnetic field detection}
\subsection{FORS\,2 observations}
\label{SECT:FORS}
Because of controversies reported in the literature about magnetic field detections and measurements performed with the FORS\,2 spectropolarimeter \citep[e.g.][]{2012A&A...538A.129B}, the data were  independently reduced in Bonn and  Potsdam using a different set  of  independent tools. The reduction and analysis performed in Bonn employed IRAF\footnote{Image Reduction and Analysis Facility (IRAF -- {\tt http://iraf.noao.edu/}) is distributed by the National Optical Astronomy Observatory, which is operated by the Association of Universities for Research in Astronomy (AURA) under cooperative agreement with the National Science Foundation.} \citep{1993ASPC...52..173T} and IDL routines based on the technique and recipes presented by \citet{2002A&A...389..191B,2012A&A...538A.129B}, while the Potsdam reduction and analysis was based on the tools described in \citet{2004A&A...415..661H,2004A&A...415..685H}, updated to include a number of statistical tests (Sch\"oller et al. 2014, in preparation). The details of the data reduction and analysis procedure applied in Bonn will be given in a separate work (Fossati et al., in preparation).

\begin{table*}[h]
\caption[ ]{Average longitudinal magnetic field values obtained from the FORS\,2 and HARPS observations.}
\label{tab:fors}
\begin{center}
\begin{footnotesize}
\begin{tabular}{l|cc|ccc|rr|rr}
\hline
 \multicolumn{10}{c}{FORS\,2}      \\
\hline
Reduction & Date & HJD$-$  & No. of  & Exp. & S/N & $\langle$B$_z\rangle$ $V$\,(G) & $\langle$B$_z\rangle$ $N$\,(G)     & $\langle$B$_z\rangle$ $V$\,(G) & $\langle$B$_z\rangle$ $N$\,(G) \\
          &      & 2450000 & frames & time\,(s) &     & \multicolumn{2}{c|}{Hydrogen}  & \multicolumn{2}{c}{All}   \\
\hline\hline

Bonn    &       \multirow{2}*{07-Feb-2014}  & \multirow{2}*{6696.7341}  & \multirow{2}*{10} &\multirow{2}*{35} &\multirow{2}*{2359}      & $-655\pm 109$ &    $22\pm 81$ &  $-504\pm  54$ &   $ 69\pm46$ \\[2pt] 
Potsdam &                                  &                            &                   &                  &                    & $-639\pm121$   & $-16\pm119$  &  $-460\pm65$   & $76\pm66$ \\[2pt] 
\hline
Bonn     & \multirow{2}*{08-Feb-2014}       &\multirow{2}*{6697.7162}   & \multirow{2}*{10} & \multirow{2}*{30}&\multirow{2}*{2398} & $-978\pm 88$ &   $-36\pm 76$ &  $-653\pm  47$ &    $40\pm  43$ \\[2pt] 
Potsdam  &  &  &  &  &  & $-877\pm91$   & $-102\pm105$  & $-521\pm62$  & $23\pm63$ \\[2pt] 
\hline
 \multicolumn{10}{c}{HARPS}      \\

\hline
Reduction & Date & HJD$-$  & No. of  & Exp.       & S/N &$\langle$B$_z\rangle$ $V$\,(G) & Detection  &  FAP&  \\
          &      & 2450000 & frames & time\,(s)  &     & \multicolumn{2}{c|}{}                            & \multicolumn{2}{c}{}   \\
\hline\hline

Bonn (LSD)    &                                  &                &    &       &         & $-592\pm7$     & DD  & $<10^{-15}$       &     \\[2pt] 
Potsdam (SVD) &        23-Apr-2014                &  6770.4993    &  4 & 2700  &   347    & $-583\pm9$     &  DD  & $<10^{-16}$       &   \\[2pt] 
Potsdam (MT)  &                                  &                &    &       &         & $-584\pm15$    &  -  & -&       \\[2pt] 
\hline
\hline

\end{tabular}
\end{footnotesize}
\end{center}
\tablefoot{Column 1 indicates the group that performed the data reduction and analysis, which led to the results shown in the following columns. The heliocentric Julian date shown in Col. 3  indicates the beginning of the sequence of exposures. Column 4 gives the number of frames obtained during each night of observation and Col. 5 the exposure time of each frame. Column 6 gives the S/N per pixel of Stokes $I$ calculated at about 4950\,\AA\ over a wavelength range of 100\,\AA. In the FORS\,2 subtable, Cols. 7 and 8 give the $\langle$B$_z\rangle$ values obtained using the spectral regions covered by the hydrogen lines obtained from the Stokes $V$ and $N$ parameter spectrum, respectively. The same is given in Cols. 9 and 10, but using the full spectrum (see Sect. \ref{SECT:FORS}). In the HARPS subtable,  Cols. 7 and 8 give the $\langle$B$_z\rangle$ $V$ values and the detection flag (DD, definite detection). Column 9 gives the false alarm probability (FAP) for those methods that provide it. The different detection techniques are listed in Col. 1 (see Sect. \ref{SECT:HARPS}).} 
\end{table*}
 

The longitudinal magnetic field ($\langle$B$_{z}\rangle$) was calculated using either the hydrogen lines or the full $3710-5870$\,\AA\ spectral region. All measurements performed on the Stokes $V$ spectrum led to detections at 6$\sigma$ or more, while we consistently got non-detections from the null profile.  Furthermore,    Monte Carlo bootstrapping tests \citep{1992nrfa.book.....P,2010MNRAS.405L..46R,2014MNRAS.440.1779H} were carried out in Potsdam. These are most often applied with the purpose of deriving robust estimates of standard errors. 
A total of 250000 tests were generated with the same size as the original dataset. The final $\langle$B$_{z}\rangle$ value was then determined from all these newly generated datasets. 
The $\langle$B$_{z}\rangle$ values and their uncertainties obtained in Bonn and Potsdam  are listed in Table \ref{tab:fors}.  The rather large difference between the $\langle$B$_{z}\rangle$ values obtained using the hydrogen lines and the entire spectrum might occur because, in presence of a strong magnetic field,  Stokes $V$ does not behave linearly with the derivative of Stokes $I$ for the metallic lines; in other words, for the metallic lines the weak-field approximation, on which the method is based, does not hold anymore. A thorough discussion about this topic can be found in \citet{2014A&A...572A.113L}.

\subsection{HARPS observations}
\label{SECT:HARPS}
As was done for the FORS\,2 data, the HARPS observations were reduced separately in Bonn and Potsdam, using independent routines. HARPS $\langle$B$_{z}\rangle$ results are also summarised in Table \ref{tab:fors}.

\subsubsection{Bonn reduction and analysis}
The reduction and analysis in Bonn was performed with the {\sc reduce} package \citep{2002A&A...385.1095P} and the Least-Squares Deconvolution technique \citep[LSD;][]{1997MNRAS.291..658D}.

The one-dimensional spectra, obtained with {\sc reduce}, were combined using the ``ratio'' method  described by \cite{2009PASP..121..993B}.  The spectra were re-normalised to the intensity of the continuum obtaining a spectrum of the Stokes $I$ ($I/I_c$) and $V$ ($V/I_c$), plus a spectrum of the diagnostic null profile \citep[$N$ - see][]{2009PASP..121..993B}, with the corresponding uncertainties. We then analysed the profiles of the Stokes $I$, $V$, and $N$ parameter using LSD, which combines line profiles (assumed to be identical) centred  at the position of the individual lines and scaled according to the line strength and sensitivity to a magnetic field (i.e. line wavelength and Land{\'e} factor). We computed the LSD profiles of the Stokes $I$, $V$, and of the null profile using the methodology and the code described in \citet{2010A&A...524A...5K}. We prepared the line mask used by the LSD code adopting the stellar parameters obtained from the spectroscopic analysis (Table \ref{StPa}). We extracted the line parameters from the Vienna Atomic Line Database \citep[{\sc vald};][]{1995A&AS..112..525P,1999A&AS..138..119K,1999PhST...83..162R} and tuned the given line strength to the observed Stokes $I$ spectrum with the aid of synthetic spectra calculated with {\sc synth3} \citep{2007pms..conf..109K}. We used all lines stronger than 10\% of the continuum, avoiding hydrogen lines, helium lines with extended wings, and lines in spectral regions affected by the presence of telluric features. The final adopted line mask contained 140 lines. We defined the magnetic field detection making use of the false alarm probability \citep[FAP;][]{1992A&A...265..669D}, considering a profile with FAP\,$<$\,$10^{-5}$ as a definite detection (DD), $10^{-5}$\,$<$\,FAP\,$<10^{-3}$ as a marginal detection (MD), and FAP\,$>$\,$10^{-3}$ as a non-detection (ND).

\begin{figure}[]
\includegraphics[width=85mm,clip]{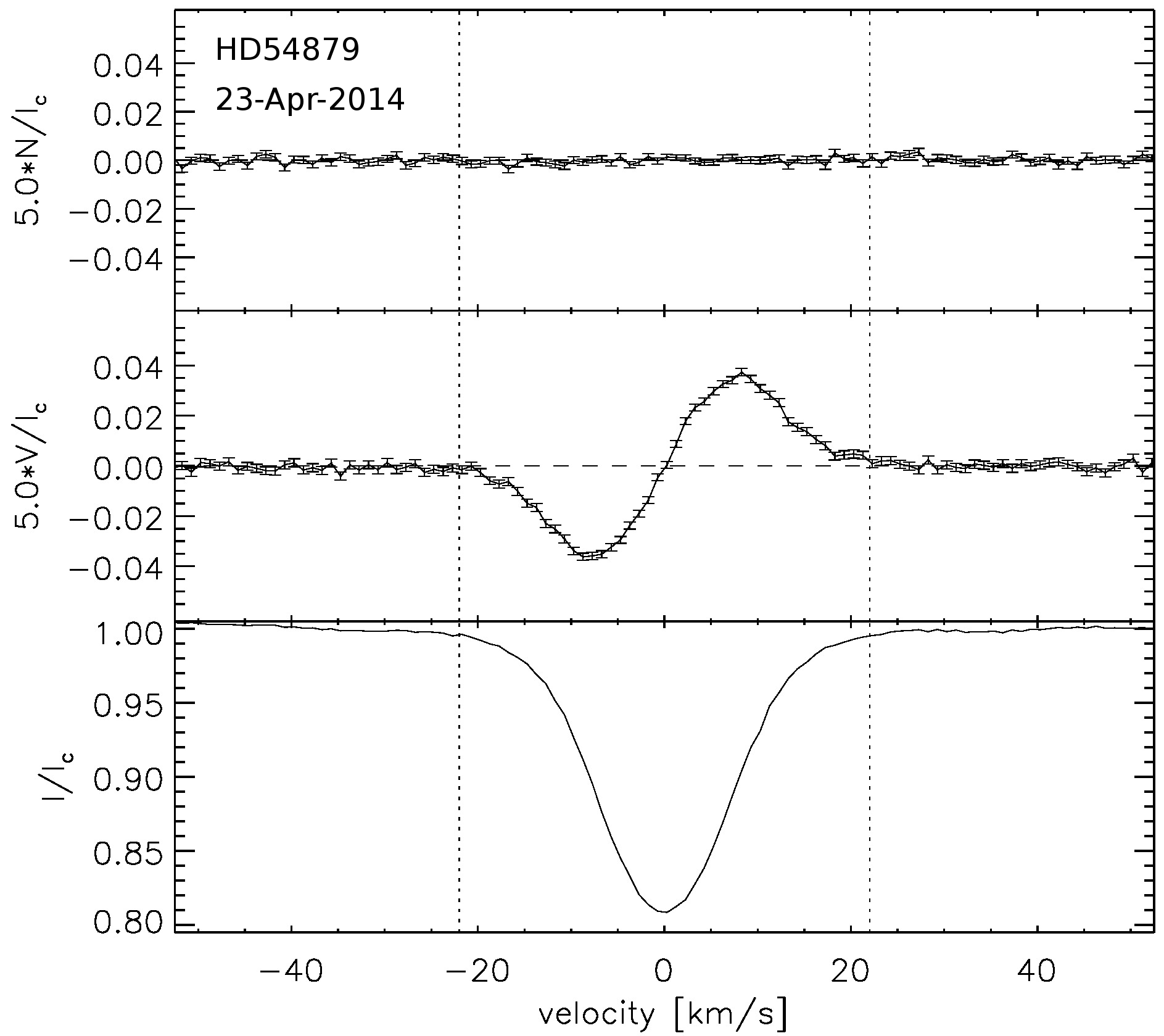}
\caption{LSD profiles of the Stokes $I$, $V$, and $N$ parameter obtained for HD\,54879. The error bars of the LSD profiles are shown for both Stokes $V$ and the null parameter. The vertical dotted lines indicate the velocity range adopted for the determination of the detection probability and magnetic field value. All profiles have been shifted upwards/downwards by arbitrary values and the Stokes $V$ and $N$ profiles have been expanded 5 times.}
\label{fig:lsd}
\end{figure}

Figure~\ref{fig:lsd} shows the LSD profiles we obtained for HD\,54879, with a S/N of the LSD Stokes $V$ profile of 3177. The analysis of the Stokes $V$ LSD profile led to a clear definite detection with a FAP\,$<$\,$10^{-15}$, while the analysis of the LSD profile of the null parameter led to a non-detection (FAP\,$>$\,0.7). Integrating over a range of 44\,$\mathrm{km\,s}^{-1}$ (i.e. $\pm$22\,$\mathrm{km\,s}^{-1}$ from the line centre) we derived $\langle$B$_z\rangle(V)$=\,$-$592$\pm$7\,G. This result  confirms the FORS\,2 magnetic field detection.

\begin{figure*}[]
\includegraphics[width=\textwidth]{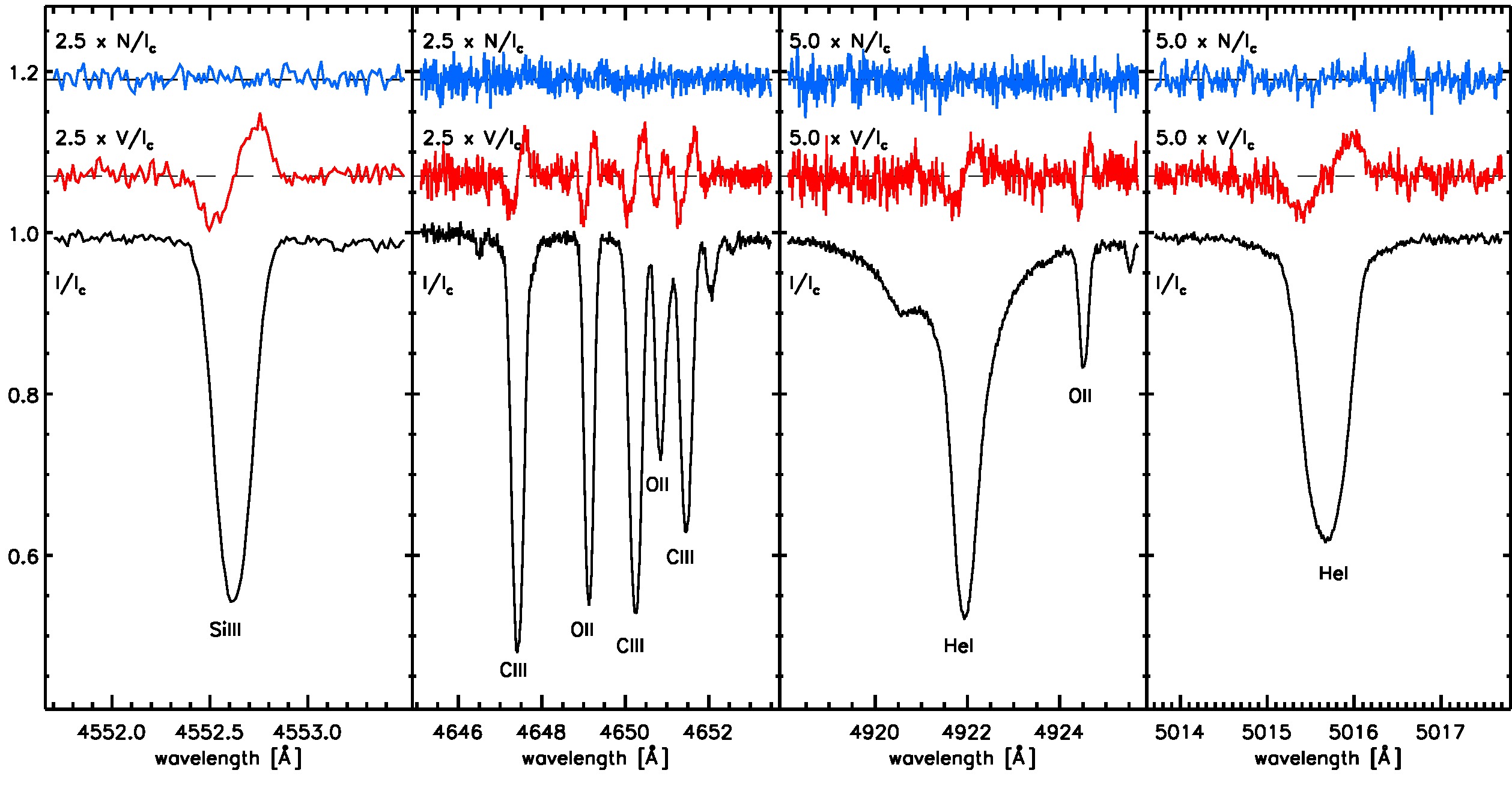}
\caption{The Stokes $I$ (black bottom line), $V$ (red middle line), and null (blue top line) profiles for a set of strong and magnetic sensitive lines. The Stokes $V$ and null profiles have been amplified by a factor of 2.5 and shifted by an arbitrary amount.}
\label{fig:multi_lines}
\end{figure*}

Because of the high S/N of the spectra and of the strong magnetic field, the Stokes $V$ profile presents an observable signature at the position corresponding to magnetic sensitive lines. Figure~\ref{fig:multi_lines} shows the Stokes $I$, $V$, and null profiles for a set of strong lines of different elements. The same shape of the Stokes $V$ profile is found for all lines. We also measured the $\langle$B$_z\rangle(V)$ value for some of them, obtaining results comparable to that given by the LSD profile.


%
\subsubsection{Potsdam reduction and analysis}

The data reduction in Potsdam  was performed with the ESO/HARPS pipeline. A detailed description of the reduction and continuum normalisation is given in \cite{2013AN....334.1093H}.

        \begin{figure}
                \resizebox{\hsize}{!}{\includegraphics[angle=0,width=\textwidth]{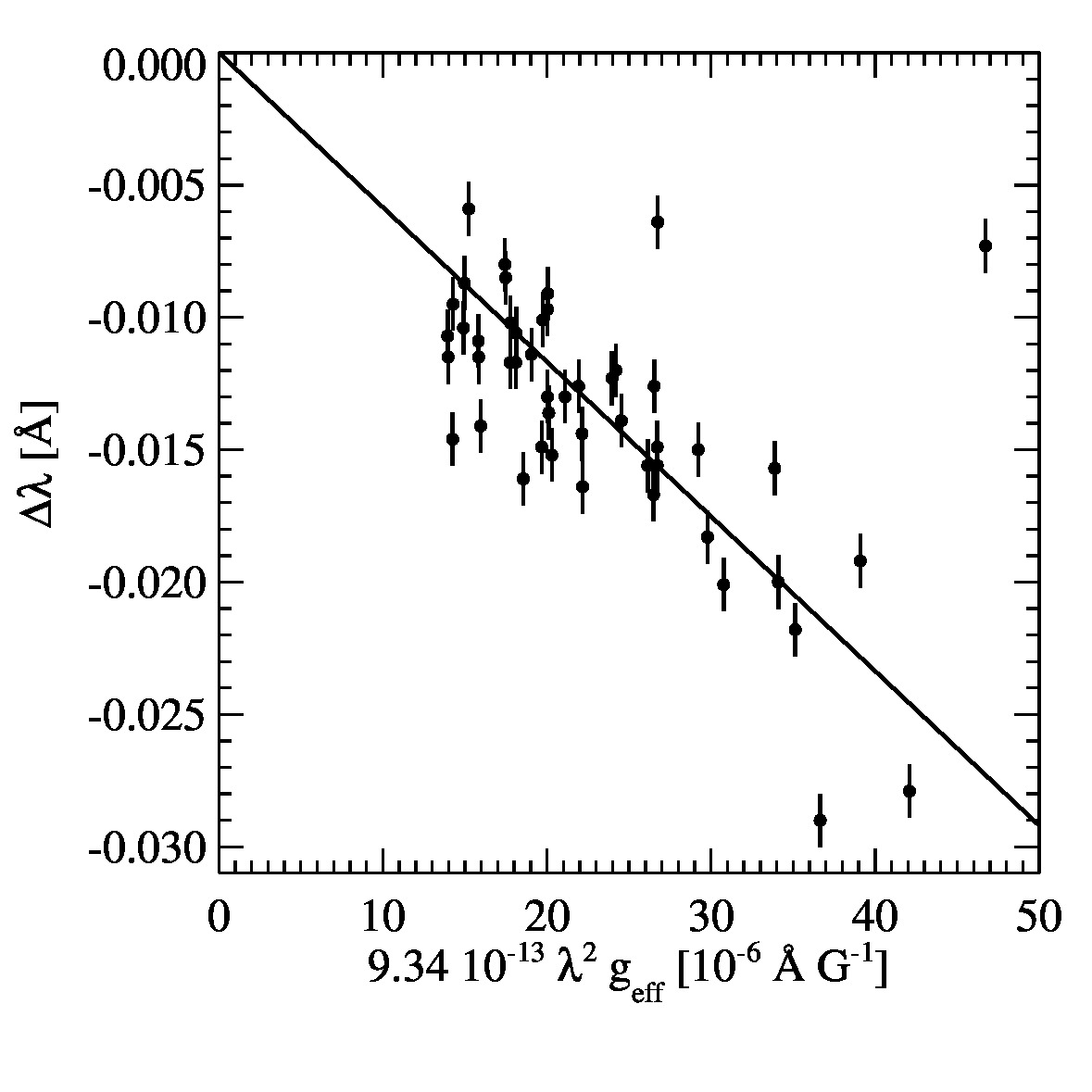}}
                \caption{ Linear regression between the displacement of each line centre of gravity in the right and left circularly polarised spectra against $9.34\,10^{-13}\lambda^2 g_{\rm eff}$ used in the moment technique \citep[][]{1991A&AS...89..121M}.  }

                \label{Fig:regr}
        \end{figure}

A total of 49, mostly unblended, metallic lines were employed in the detection of a surface magnetic field using the moment technique \citep[MT,][]{1991A&AS...89..121M}. This technique allows us to determine the mean longitudinal field strength and demonstrate the presence of the crossover effect and quadratic magnetic fields, and so to constrain the magnetic field topology in more detail than can be done with the LSD and Singular Value Decomposition (SVD) methods. For each line in the sample of metallic lines, the measured shifts between the line profiles in the left- and right-hand circularly polarised HARPS spectra are used in a linear regression analysis in the $\Delta\lambda$ versus $\lambda^2 g_{\rm eff}$ diagram, following the formalism discussed by \cite{1991A&AS...89..121M,1994A&AS..108..547M}  (see also \citealt{2014arXiv1406.1927H}, Fig. 11).  The corresponding linear regression is shown in Fig.~\ref{Fig:regr}. The Land\'e factors were taken from Kurucz's list of atomic data\footnote{http://kurucz.cfa.harvard.edu/atoms}. For each line measured, the mean error was calculated taking into account the signal-to-noise of the spectra and the uncertainty of the wavelength calibration \citep{1994A&AS..108..547M}. We obtained  a mean longitudinal magnetic field of $\langle$B$_z\rangle(V)$=\,$-$584$\pm$15\,G.  In addition, we derived a $\langle$B$_z\rangle(N)$=\,$-$22$\pm$10\,G from the spectrum of the $N$ parameter, calculated by combining the subexposures in such a way that polarisation cancels out.  Since no significant  magnetic field could be detected from the null spectrum, we concluded that any noticeable spurious polarisation is absent. This is confirmed by the analysis of the LSD $N$ profile. No crossover and mean quadratic magnetic field have been detected for this observational epoch.

\begin{figure}[]
\includegraphics[width=90mm,clip]{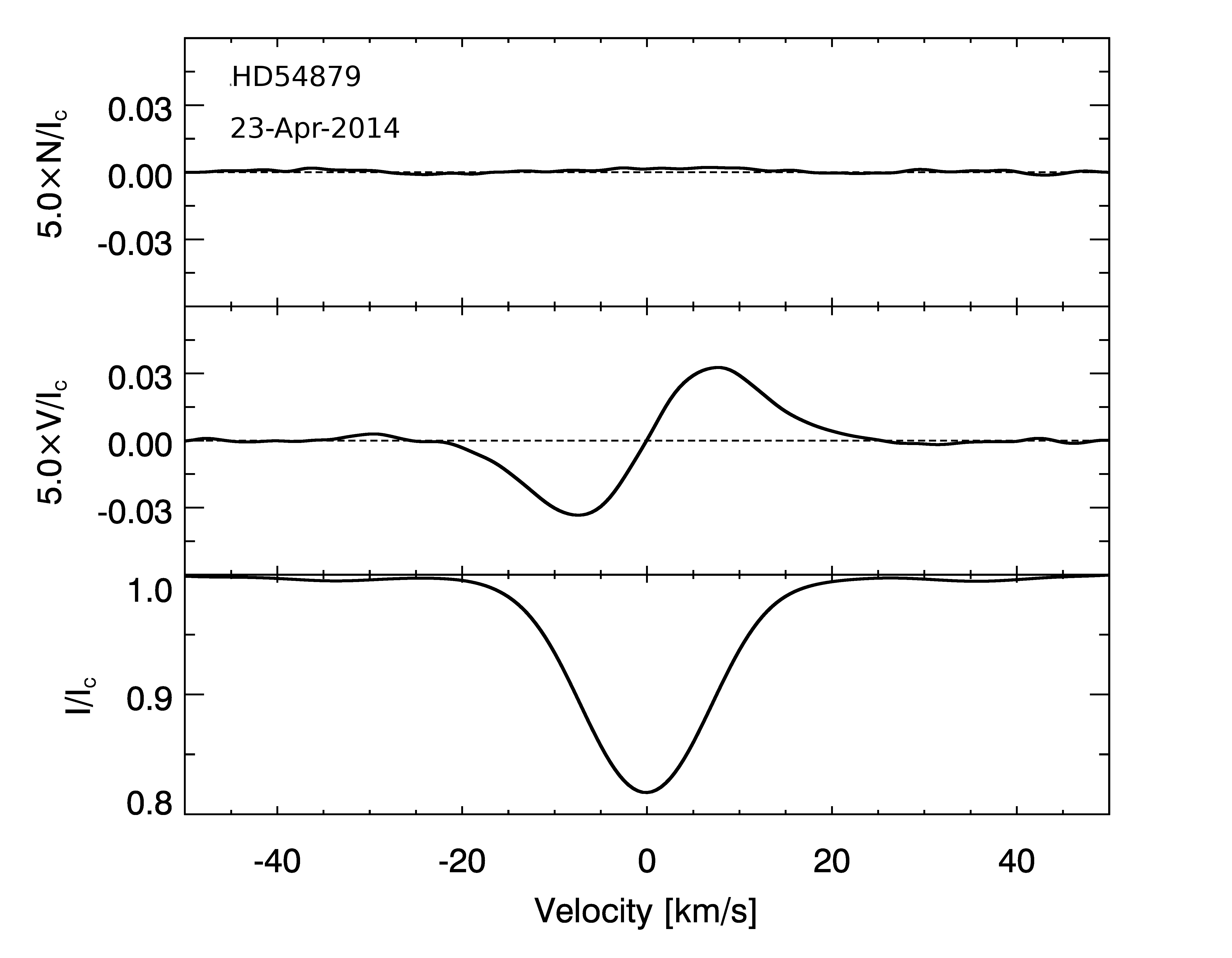}
\caption{SVD profile of Stokes $I$,  $V$, and of the Null profile $N$. For display purposes the SVD profiles have been corrected for the radial velocity determined from the Stokes $I$ profile. Like the LSD profiles, the SVD Stokes $V$ and the $N$ profiles have been  shifted by an arbitrary amount and expanded by a factor of 5.}
\label{fig:svd}
\end{figure}

In addition to the moment technique, the magnetic field was  detected using the multi-line SVD technique for Stokes Profile Reconstruction, recently introduced by \cite{2012A&A...548A..95C}. The results obtained with the SVD  technique, using a mask of 160 lines that excludes helium and hydrogen lines, are  shown in Fig \ref{fig:svd}.  As was done for the LSD analysis, the line mask was constructed using {\sc vald}. The  analysis led to a definite detection with a FAP smaller than 10$^{-16}$. From the retrieved Stokes $V$ profile we derived a mean longitudinal magnetic field  $\langle$B$_z\rangle(V)$=\,$-$583$\pm$9\,G.

\section{Stellar parameters and abundances}                
\label{HIGHLOW}

The observations presented in Sect.~\ref{Obs} provided us with data at low and high spectral resolution, both datasets having a high S/N ($>$\,$300$). We present the quantitative analysis of the datasets, aiming at 1) characterising HD\,54879 through  the spectra of two different instruments (i.e. FORS\,2 and HARPS) and 2) exploring the impact of spectral resolution on the stellar parameters and chemical abundances.

We measured the  \ensuremath{{\upsilon}\sin i} value of nearly 130 metal transitions from the HARPS spectrum using the tool {\sc iacob-broad} developed by \cite{2014A&A...562A.135S}. 
We obtained an average \ensuremath{{\upsilon}\sin i} value of $7\pm2$\,km\,s$^{-1}$ and a macroturbulence of $8\pm3$\,km\,s$^{-1}$, in agreement with \cite{2014A&A...562A.135S}.  We adopted these values for our  analyses.

The quantitative atmosphere analysis was performed using the stellar atmosphere code {\sc fastwind} \citep[][]{1997A&A...323..488S,2005A&A...435..669P}. The code enables non-LTE calculations and assumes a spherical symmetry. The velocity structure of the stellar wind is modelled with a $\beta$ velocity law. Following the technique described by \cite{castro2011} (see also \citealt{2005ApJ...622..862U} and \citealt{Lefever_2007}), the stellar parameters and chemical abundances were derived through automatic algorithms searching for the  set of parameters that best reproduce the main transitions in the range $\sim$\,$4000-4900$\,\AA. The automatic tools were based on a large collection of {\sc fastwind} stellar grids (\citealt{castro2011,2011JPhCS.328a2021S}). The chemical analysis procedure has been  updated since \cite{castro2011} by implementing an optimised genetic algorithm.

The best fitting parameters obtained from the analysis of the FORS\,2 and HARPS spectra are listed in Table~\ref{StPa}, where for comparison we added the stellar parameters of the O9.7~V standard  HD\,36512 \citep{2011ApJS..193...24S}, obtained by S. Sim\'on-D\'iaz using the {\sc iacob-gbat} code \citep{2011JPhCS.328a2021S}. The observed FORS\,2 and HARPS spectra and the best-fit models are plotted in Figs. \ref{Fig:FORS} and \ref{Fig:HARPS},  see also Figs. \ref{Fig:HARPSA1}\,--\,\ref{Fig:HARPSA3} for a  detailed display of the HARPS data. The lines used in the analyses  are  marked. Table \ref{StPa} shows the good agreement between the stellar parameters derived from  the analysis of the high- and low-resolution spectra.  In addition, the HD\,54879  temperature and gravity values agree with those obtained for the standard star HD\,36512.  The analysis of the FORS\,2 and HARPS spectra yield  $\rm{He/H}\sim0.10-0.12$. The low sensitivity of the He lines to abundance variations prevents us from obtaining a more accurate value. Nevertheless, we exclude a substantial helium enhancement/depletion for this star. 

The  wind-strength Q-parameter\footnote{The Q-parameter \citep{1996A&A...305..171P,2000ARA&A..38..613K} is a measurement of the stellar wind strength defined as $Q = \dot{M} \, / \, \left(R_{*}v_{\infty}\right)^{1.5}$, where $\dot{M}$ is the mass-loss rate, $R_{*}$ the stellar radius, and $v_{\infty}$ the terminal wind velocity.}  estimated  in HD\,54879, mainly based on the Balmer lines, is substantially  higher (log\,$Q=-11.0$) than expected for a late-O dwarf (i.e. log\,$Q=-13.46$ in HD\,36512). To further investigate this difference, Fig.~\ref{Fig:STD} compares high-resolution spectra of HD\,54879 and HD\,36512. Both spectra were taken with the same instrument as part of the IACOB project (see Sect. \ref{var}).  Convolving the spectrum of HD\,54879 to match the rotation and macroturbulence velocities of HD\,36512 reveals a remarkable similarity between the spectra of the two stars. However,  the transitions formed in the outer parts of the atmosphere show noticeable differences, especially H$\alpha$.  It is unlikely that an exceptionally strong stellar wind is at the origin of the H$\alpha$ emission, which instead  most probably originates from circumstellar material in the magnetosphere (see Sect. \ref{magnetic}), as  already reported in other stars \citep[e.g.][]{1980ApJS...44..535W,2009MNRAS.400L..94G}. We therefore re-calculated the stellar parameters without considering the lines most sensitive to the  external atmospheric layers. The new parameters obtained in this way are identical to the  original values.

\begin{table*}[!th]
\centering
\caption{Stellar parameters and chemical abundances of HD\,54879 and of the O9.7~V standard star HD\,36512 \citep{2011ApJS..193...24S}.}

\begin{tabular}{l l r r r r r}
\hline
ID & Instrument &   $T_{\rm{eff}}$  &  log$\,g$ & log$\,Q$  &  $\xi^{\,b)}$& He/H  \\  
&     &  (K)              &   (dex)  &     &  (km\,s$^{-1}$)  & (by number)    \\
\hline
\hline\\
\multirow{2}{*}{HD\,54879}&     FORS\,2  & 33000 $\pm$  1000 & 3.95 $\pm$ 0.10 & \textit{-}$^{\,a)}$  & 5 $\pm$ 3 & $0.10-0.12$ \\[2pt] 
&       HARPS & 33000 $\pm$  1000 & 4.00 $\pm$ 0.10 & \textit{-}$^{\,a)}$  & 4 $\pm$ 1 & $0.10-0.12$ \\[2pt] 
HD\,36512&      FIES & 33400 $\pm$  600\hspace{0.175cm} & 4.09 $\pm$ 0.11 & -13.46  & 10  & $0.10$ \\[2pt] 
\hline
&       &log\,$\epsilon_{\ion{Si}{}}$&  log\,$\epsilon_{\ion{Mg}{}}$&   log\,$\epsilon_{\ion{C}{}}$&    log\,$\epsilon_{\ion{N}{}}$&    log\,$\epsilon_{\ion{O}{}}$      \\
\hline\\
\multirow{2}{*}{HD\,54879} &    FORS\,2   & $7.3\pm0.3 $ & $7.4\pm0.3$ & $7.8\pm0.3  $ & $7.5\pm0.3$ & $8.6\pm0.2$ \\[4pt]
&       HARPS  & $7.4\pm0.2$ &  $7.4\pm0.1$ & $8.1\pm0.2  $ & $7.7\pm0.2$ & $8.6\pm0.1$ \\[4pt]

\hline

   $\odot$ & &   7.51     &7.60  &   8.43        &   7.83        &   8.69                    \\
        CAS     & &   $7.50$     & $7.56$  &   $8.33$      &   $7.79$        &   $8.76$                    \\

\hline\hline

\end{tabular}
\tablefoot{$a)$  The parameter $Q = \dot{M} \, / \, \left(R_{*}v_{\infty}\right)^{1.5}$ \citep{1996A&A...305..171P} relies mainly on the Balmer lines, which for HD\,54879 are not suitable for setting the mass-loss rate (Sect.~\ref{HIGHLOW}). $b)$ Microturbulence. The solar abundances ($\odot$) are taken from \cite{2009ARA&A..47..481A}. The cosmic abundance standard  (CAS) composition from \cite{2012A&A...539A.143N} is also listed, log\,$\epsilon_{\ion{X}{}}=\,$log$\,(\ion{X}{}/\ion{H}{})+12$ (by number).}
\label{StPa}
\end{table*}


We used \textsc{bonnsai}\footnote{The \textsc{bonnsai} web-service is available at http://www.astro.uni-bonn.de/stars/bonnsai} \citep{Schneider+2014c} to determine the current  mass, radius, and age of HD\,54879. \textsc{Bonnsai} computes the full posterior probability distributions of stellar parameters using Bayes' theorem. The code simultaneously matches  the derived effective temperature, surface gravity, and projected rotational velocity of HD\,54879 to the Milky Way single-star models of \citet{2011A&A...530A.115B}. We assumed a Salpeter initial mass function \citep{1955ApJ...121..161S} as initial mass prior and uniform priors for the age and initial rotational velocity. The stellar rotation axes are randomly oriented in space. 

\begin{figure}
                \resizebox{\hsize}{!}{\includegraphics[angle=0,width=\textwidth]{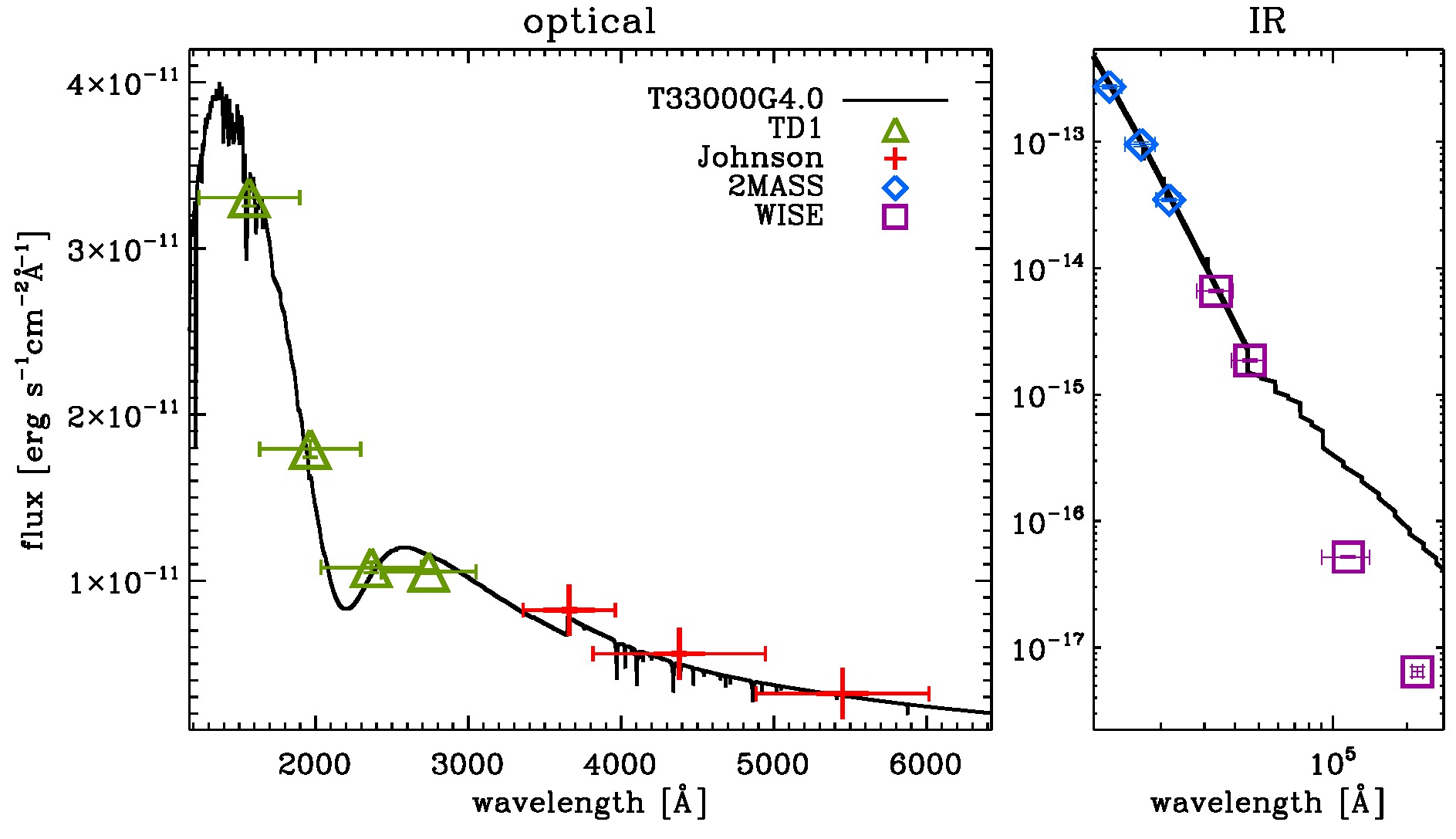}}
                \caption{ Comparison between {\sc fastwind} (full black line) synthetic flux and TD1 (green triangles), Johnson (red pluses), 2MASS (blue diamonds), and WISE (purple squares) photometry converted into physical units. The UV and optical spectral regions are shown in the left panel and  the IR band in the right panel. }

                \label{Fig:SED}
        \end{figure}

\begin{figure*}
                \resizebox{\hsize}{!}{\includegraphics[angle=90,width=\textwidth]{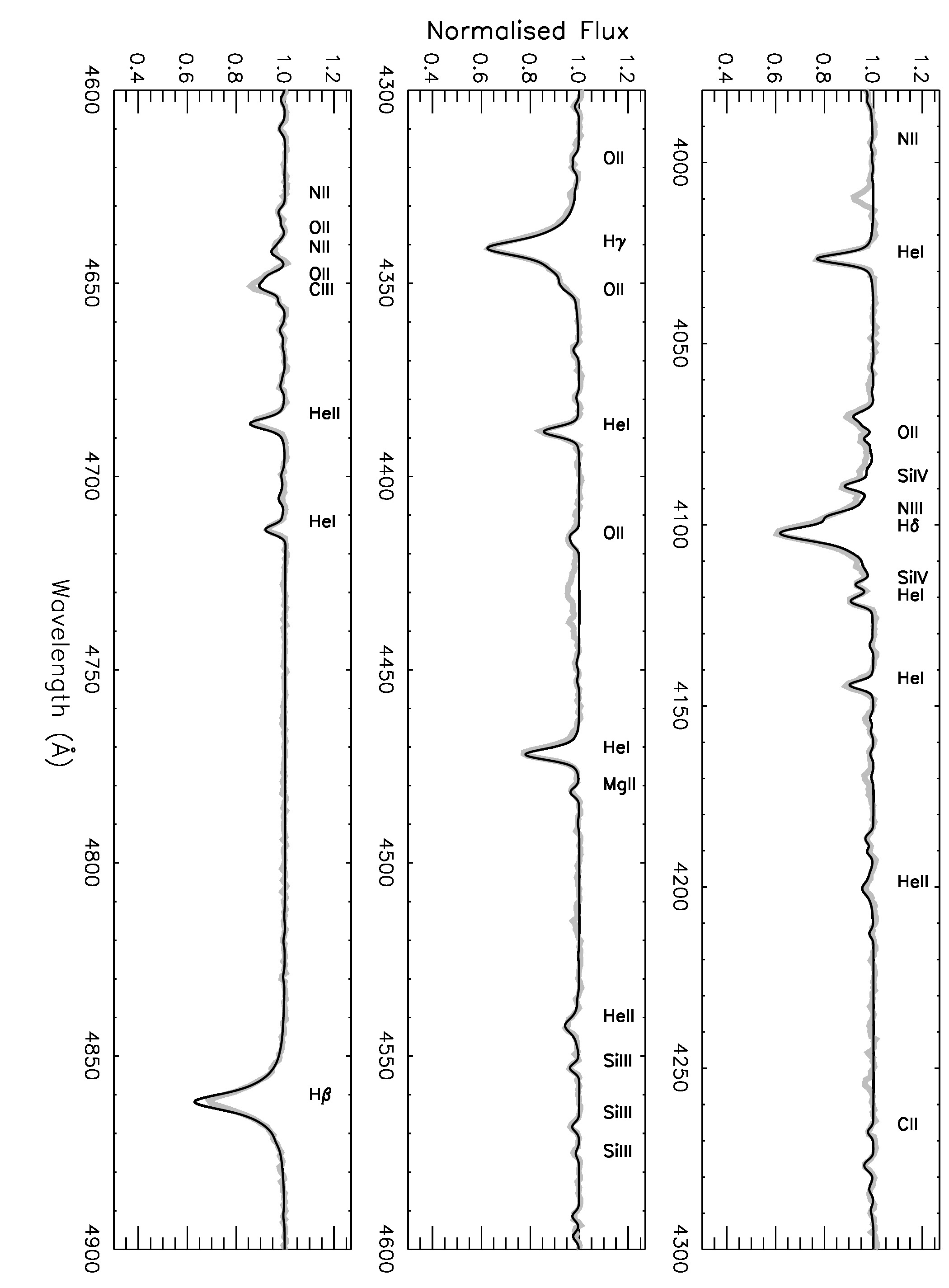}}
                \caption{Normalised FORS\,2 optical spectrum (grey) and the best-fit \textsc{fastwind} stellar model (black). The lines included in the analysis are marked.}
                \label{Fig:FORS}
        \end{figure*}

\begin{figure*}
                \resizebox{\hsize}{!}{\includegraphics[angle=90,width=\textwidth]{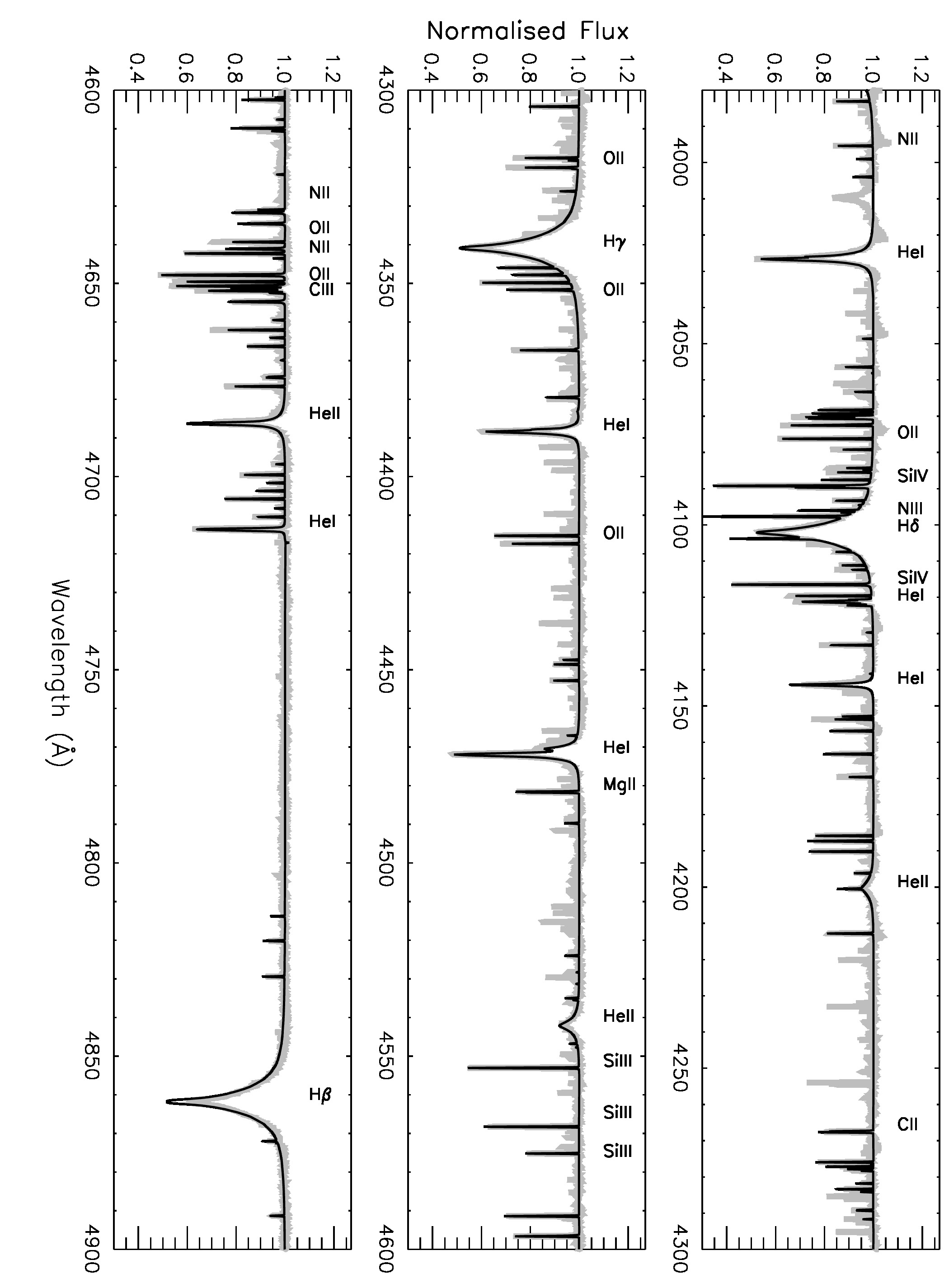}}
                \caption{Same as Fig. \ref{Fig:FORS} but for the HARPS normalised spectrum. Only the main transitions for the spectral and chemical analysis have been modelled \citep{castro2011}.}
                \label{Fig:HARPS}
        \end{figure*}

\begin{figure*}
                \resizebox{\hsize}{!}{\includegraphics[angle=90,width=\textwidth]{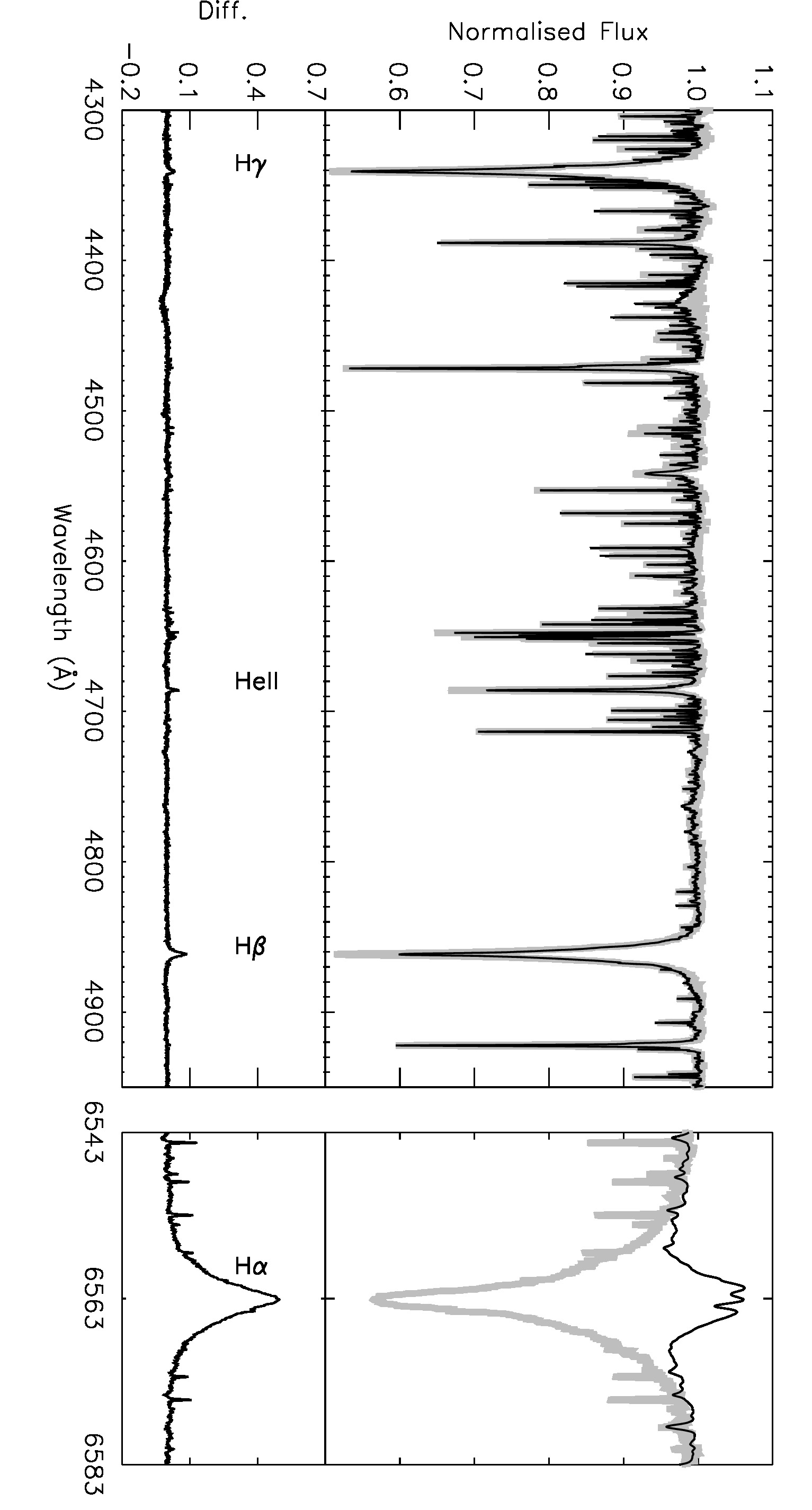}}
                \caption{Top panels:  optical spectra of HD\,54879 taken on Jan. 15, 2011 (black), and the O9.7~V standard HD\,36512 (grey). Bottom panels: difference between the two stars, after convolving HD\,54879 to match the broadening of HD\,36512. The lines showing the most substantial differences are highlighted. The mismatch at 4428\,\AA\ is the result  of a diffuse interstellar band. }
                \label{Fig:STD}
        \end{figure*}

The fundamental parameters derived by {\sc bonnsai} are listed in Table \ref{BONNSAI} together with the spectroscopic values. The spectroscopic luminosity ($L$), current mass ($M$), and radius ($R$) were calculated taking a distance of 1.32\,kpc \citep{1978ApJS...38..309H}, the spectral energy distribution provided by  {\sc fastwind}, the extinction laws of \cite{2007ApJ...663..320F}, and HD\,54879 optical and 2MASS photometry.  Figure \ref{Fig:SED} shows the comparison of the {\sc fastwind} synthetic fluxes, calculated by adopting the fundamental parameters derived for HD\,54879 with the available ultraviolet \citep[UV,][]{1978csuf.book.....T}, optical \citep{1994cmud.book.....M}, and infrared \citep[IR,][]{2003yCat.2246....0C,2012yCat.2311....0C} photometry. The discrepancy between the synthetic flux and WISE photometry is a known issue \citep{2014A&A...562A.143F}. WISE magnitudes were not used in the  spectroscopic parameters reported in Table \ref{BONNSAI}. 
The match between {\sc bonnsai} and the spectroscopic values highlights the consistency between the stellar tracks prediction, distance, and photometry of HD\,54879.  The fractional main-sequence age derived from the \cite{2011A&A...530A.115B} evolutionary tracks is 0.46. The age derived with {\sc bonnsai} (see Table \ref{BONNSAI}) agrees with that of the CMa OB1 association of which the star is a probable member \citep[3\,Myr,][]{1974A&A....37..229C}.

The  obtained non-LTE chemical abundances are listed in Table \ref{StPa}, together with the solar abundances by \cite{2009ARA&A..47..481A} and the present-day massive star abundances in the solar neighbourhood  by \cite{2012A&A...539A.143N}. The abundance values derived from the HARPS and FORS\,2 spectra agree within the errors. The low resolution hampers a precise abundance determination of some chemical elements. For instance, the measurement of the nitrogen abundance is based on weak lines (e.g.  N\,{\sc ii} $\lambda\sim$3995\,\AA) that are clearly visible in the HARPS spectrum (Fig.~\ref{Fig:HARPS}), but blurred in the continuum at the low resolution of FORS\,2 (Fig.~\ref{Fig:FORS}). Oxygen presents strong lines that are clearly visible, hence measurable, even at low resolution.

\begin{table}
\centering
\caption{HD\,54879 theoretical ({\sc bonnsai}) and spectroscopic (Spec.)  stellar parameters.} 
\begin{tabular}{ l r r}
\hline
HD\,54879&       {\sc bonnsai}   & Spec.         \\
\hline\hline\\
log\,$L/L_\odot$   & $4.7^{+0.2}  _{-0.2}$   & $4.7^{+0.3}  _{-0.2}$  \\[4pt]
$R/R_\odot$        & $6.7^{+1.0}  _{-0.9}$   & $6.8^{+2.3}  _{-1.6}$   \\[4pt]
$M/M_\odot$        & $18.6^{+2.0} _{-1.6}$   & $16.9^{+1.1} _{-1.0}$   \\[4pt]
Age (Myr)         & $4.0^{+0.8}  _{-1.2}$  & \\[4pt]
\hline
\hline

\end{tabular}
\label{BONNSAI}
\end{table}


\section{Discussion}           

\subsection{General considerations about the detected magnetic field}
\label{magnetic}

The  $|\langle$B$_z\rangle|$ values independently obtained in Bonn and Potsdam agree within the uncertainties.  The FORS\,2 results from Bonn indicate slightly larger fields than those from Potsdam. The magnetic fields derived from HARPS data in Bonn and Potsdam are practically identical. The LSD, SVD, and MT, on average, provide $\langle$B$_z\rangle(V)$=$-586\pm5$\,G.

 With the few available magnetic field measurements it is not possible to perform a meaningful modelling of the magnetic field topology and strength, particularly with measurements conducted with completely different instruments and techniques \citep{2014A&A...572A.113L}. Nevertheless, from the maximum recorded $|\langle$B$_z\rangle|$ value it is possible to derive the expected minimum dipolar magnetic field strength \citep[B$_d$; e.g. Eq 7 of][]{2007A&A...475.1053A}. From both FORS\,2 and HARPS measurements we derived a maximum $|\langle$B$_z\rangle|$ of the order of 600\,G, implying B$_d$\,$\gtrsim$\,2.0\,kG. Despite the strong magnetic field, we have found no sign of photospheric line profile variations (Sect.~\ref{var}).

The dipolar field strength of HD\,54879 places this star at the same magnetic strength level as the fast-rotating secondary of Plaskett's star \citep{2013MNRAS.428.1686G} and ALS\,15218 \citep{2012MNRAS.423.3413N}. Radial velocity variations have been reported for ALS\,15218 \citep{2011BSRSL..80..644C}, so it may also be part of a binary system. In contrast, HD\,54879 is apparently a slowly rotating single  star, though we cannot rule out a pole-on view, and is probably the  strongest magnetic non-peculiar and single O-type star detected so far.

In the context of the classification of magnetospheres of massive stars presented by \citet{2013MNRAS.429..398P} and assuming a minimum dipolar magnetic field strength of 2.0\,kG and an equator-on view, we obtained a lower limit on the Alfv{\'e}n radius of about 5 stellar radii and an upper limit on the Keplerian corotation radius of about 20 stellar radii. For the calculation of the Alfv{\'e}n and Keplerian corotation radius we adopted the stellar parameters obtained from {\sc bonnsai}, a terminal velocity of 1700\,$\mathrm{km\,s}^{-1}$ \citep{2000ARA&A..38..613K} and the mass-loss rate obtained from the relation given by \citet{2000A&A...362..295V}. The derived values indicate that the star has a dynamical magnetosphere, but one has to keep in mind that the Alfv{\'e}n radius is a lower limit and the Keplerian corotation radius is an upper limit, hence the star could have a centrifugal magnetosphere \citep{2005MNRAS.357..251T,2009MNRAS.394..415M,2013MNRAS.429..398P}. Moreover, HD\,54879 is in the weak wind regime of late-O dwarfs as discussed by \cite{2005A&A...441..735M}. In this regime, the mass-loss rates might be up to a factor of hundred lower than expected from theory  \citep{2008A&ARv..16..209P}, whereas the hydrodynamically measured rate of the O9.5 dwarf $\zeta$\,Oph \citep{2012MNRAS.427L..50G} is found to be only a factor of six below that predicted by \citet{2000A&A...362..295V} \citep[see also][]{2012ApJ...756L..34H}. A smaller mass-loss rate would lead to a larger Alfv{\'e}n radius, placing the star closer to the region covered by stars with a centrifugal magnetosphere; this region could then be reached with a slightly shorter rotational period (i.e., a non-orthogonal inclination angle).

The mass-loss rate obtained for HD\,54879, assuming a line-driven stellar wind as the unique cause of H$\alpha$  emission, is too large according to its spectral type and previous studies of late O-type dwarfs \citep[Fig. \ref{Fig:STD}, see also][]{2006A&A...448..351S,2011A&A...535A..32N,2012A&A...538A..39M}. HD\,54879 fits in the Oe-star category, as defined by \cite{1974ApJ...193..113C} \citep[see also][]{2004AN....325..749N}. The classical Be scenario suggests a  circumstellar disk produced by a  rapidly  rotating star  \citep[][]{2003PASP..115.1153P}. In the case of HD\,54879, a centrifugal magnetosphere supported by the strong magnetic field could be responsible for the circumstellar H$\alpha$ emission.

 The low \ensuremath{{\upsilon}\sin i} observed in HD\,54879 could also be a consequence of the strong magnetic field \citep{1958HDP....51..689D,2013MNRAS.433.2497S}. Magnetic braking can remove angular momentum when the stellar outflow remains coupled to the magnetic field as it leaves the star \citep[e.g.][]{1987MNRAS.226...57M,2009MNRAS.392.1022U,2013MNRAS.429..398P}.

\subsection{General considerations about the stellar atmosphere analysis}

The analysis of the stellar parameters places HD\,54879 on the main-sequence phase with an age of $4.0^{+0.8}  _{-1.2}$\,Myr, according to the \cite{2011A&A...530A.115B} evolutionary tracks. The chemical abundances, inferred from the HARPS data, are slightly lower than the solar values \citep{2009ARA&A..47..481A} and the cosmic abundance standard obtained  by \cite{2012A&A...539A.143N}, though still compatible within the uncertainties. Neither helium nor the other  chemical elements analysed show any noteworthy peculiarity.

The differences between spectroscopic and evolutionary masses have been a source of conflict  \citep{1992A&A...261..209H}, although improvements in both fields have reduced the discrepancies \citep{2007A&A...465.1003M,2012A&A...543A..95R,2013A&A...555A...1B}. HD\,54879 shows a difference in log$\,M_{\rm{spec}}/M_{\rm{evol}}=\,-0.04\pm0.05\,$dex. Although this is only one star, the result supports the reliability of our routines, analysis techniques, and stellar models. Nevertheless, a large sample of stars is required to determine whether the systematic log$\,M_{\rm{spec}}/M_{\rm{evol}}=-0.06\,$dex offset reported by \cite{2009ApJ...704.1120U} \citep[see also Fig. 9 in][]{castro2011} still persists when taking into account the latest state-of-the-art  developments in the stellar theory and analysis techniques \citep[see also][]{2014arXiv1409.7784M}.

We would like to note the good agreement between the high and low spectral resolution stellar atmosphere analyses. This test provides additional confidence in previous studies carried out using low spectral resolution  data \citep[e.g.][]{2007ApJ...659.1198E,2009ApJ...704.1120U,castro2011}, with a note of caution concerning the chemical analysis   described in Sect. \ref{HIGHLOW}.


\subsection{Radial velocities and line profile variations}
\label{var}
The strong magnetic field detected in HD\,54879 could lead to line profile variations \citep[see e.g.][]{2002A&A...381..736P,2013A&A...554A..93K}. To check for spectral variability we have complemented our HARPS observations with high-resolution (R\,$\sim$\,$50000$) spectra from the IACOB \citep{2011IAUS..272..310S,2014A&A...562A.135S,2015arXiv150404257S} and OWN \citep{2010RMxAC..38...30B} surveys. This resulted in eight additional spectra (three obtained with FIES@NOT2.5 and  five with FEROS@ESO2.2).   All the collected high-resolution spectra were obtained between 2009 and 2014. The observing dates and radial velocity measurements, derived from the average of the individual lines marked in Fig. \ref{Fig:HARPS}, are summarised in Table \ref{vra}. \cite{2007PASP..119..742B} calculated for HD\,54879 a constant radial velocity of $35.4\pm1.4\,$km\,s$^{-1}$. The authors also mentioned a previous radial velocity measurement by \cite{1943ApJ....97..300N} of $15.6\pm1.4\,$km\,s$^{-1}$.  These results  might suggest that the star is a member of a long-period (i.e. tens of years) binary, although no companion was found in interferometry studies  by \cite{1998AJ....115..821M} and  \cite{2014arXiv1409.6304S}. Given the negligible radial velocity variation measured from the spectra listed in Table \ref{vra}, and a lack of companions in high angular resolution studies, we assume the single-star scenario for HD\,54879.

\begin{figure*}
                \begin{subfigure}{0.5\textwidth}                
                \includegraphics[width=\textwidth]{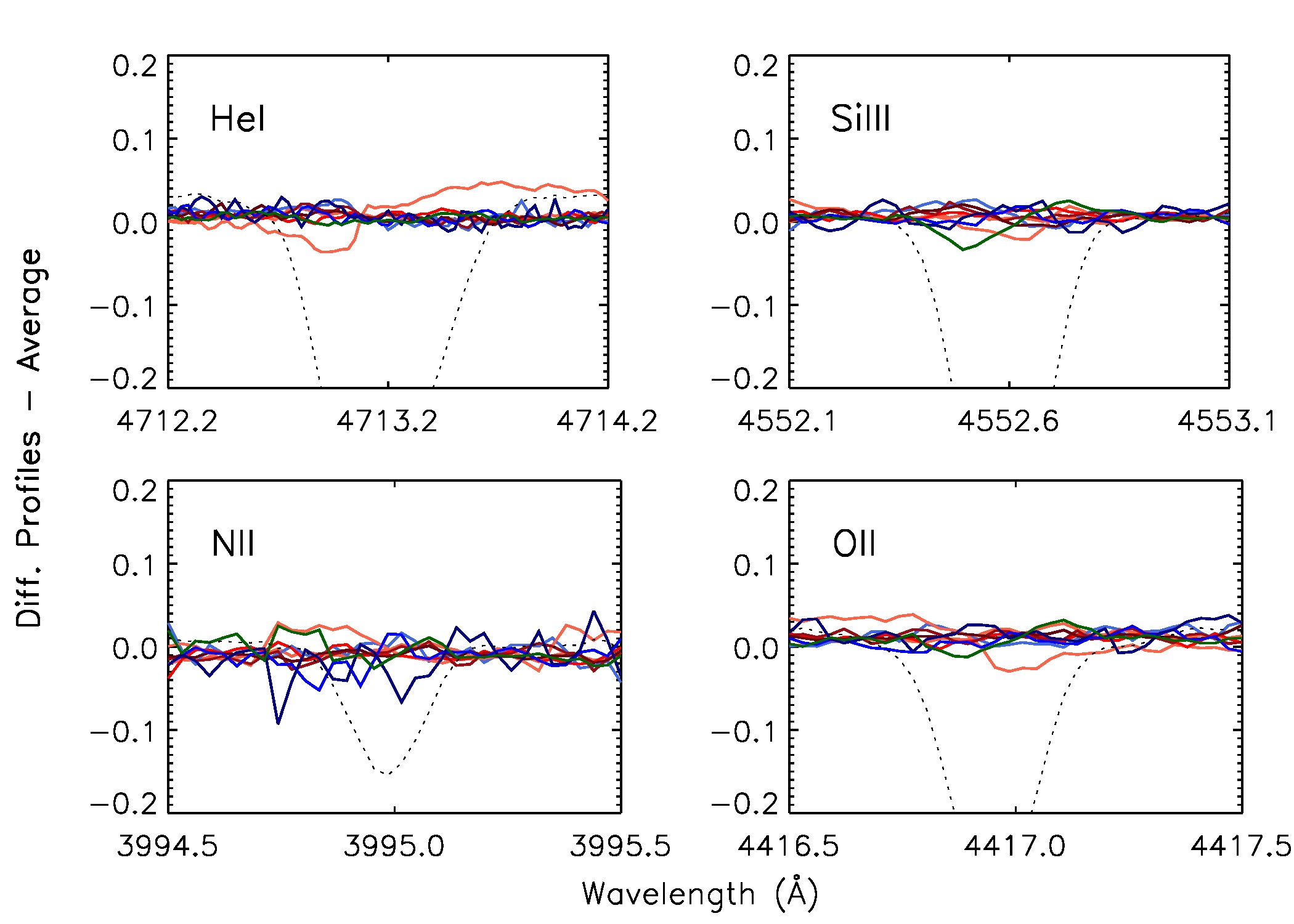}
                 \end{subfigure}%
                \begin{subfigure}{0.5\textwidth}                
                \includegraphics[width=\textwidth]{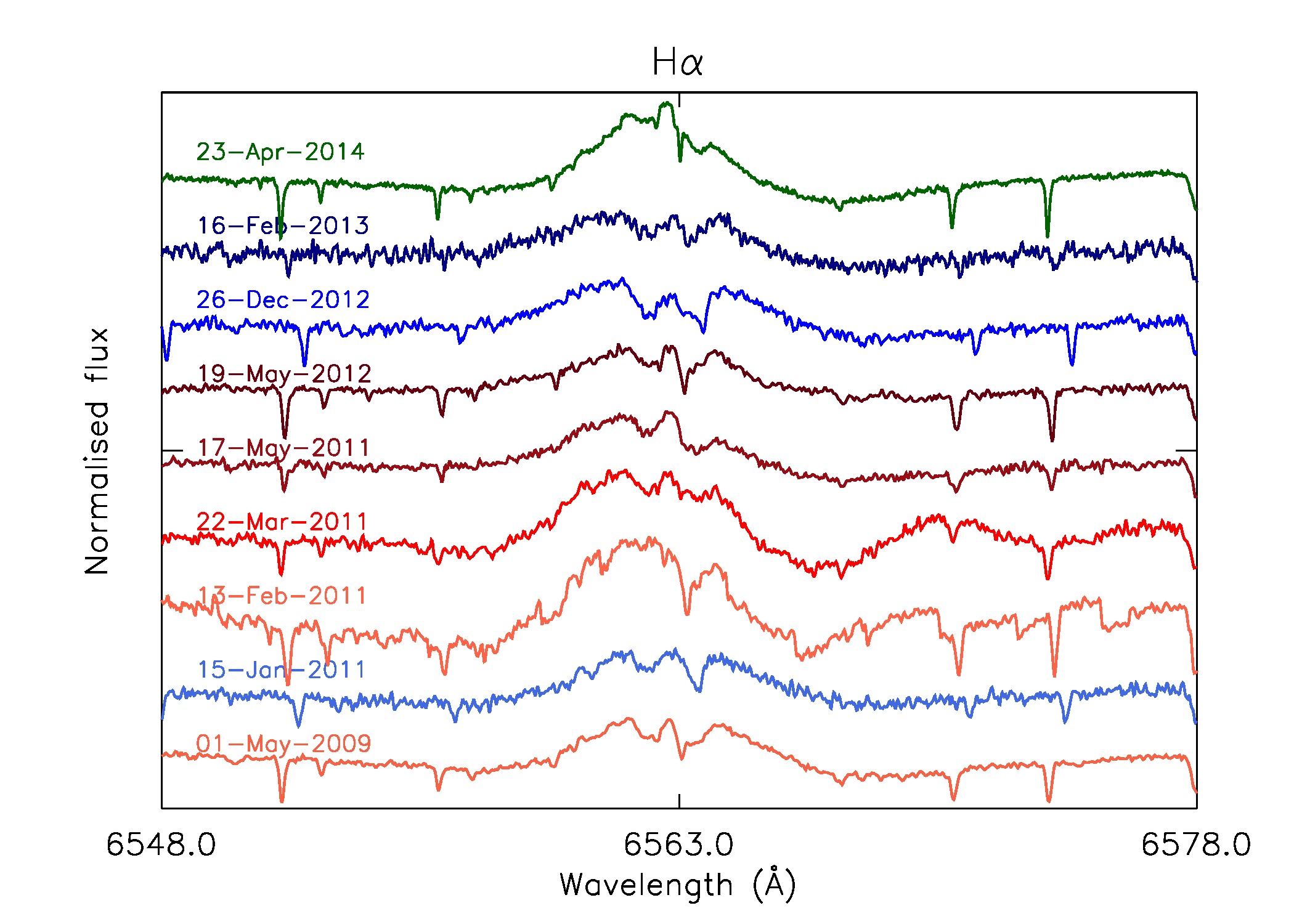}
                 \end{subfigure}%
                \caption{Left panels: difference for four lines between the average profile and all the observations presented in Table \ref{vra}. The colour code is indicated in the right panel. The average profiles are overplotted  (black dotted line). Right panel: Time series of H$\alpha$ profile between 2009 and 2014. The narrow absorption transitions correspond to telluric features.}
                \label{Fig:FEROS}
                
        \end{figure*}

By combining the stellar radius (Table \ref{BONNSAI}) and the \ensuremath{{\upsilon}\sin i} value obtained for HD\,54879, we derived a maximum  rotation period of 43 days. Given the short period and time sampling of our high-resolution spectra, line profile variations due to spots should be clearly detectable. The left panel of Fig. \ref{Fig:FEROS} reveals instead that the shape of the photospheric lines does not vary with time. A pole-on view of HD\,54879 could explain the lack of line profile variations, but the magnetic field variations discard this hypothesis. In contrast, the H$\alpha$ line (right panel of Fig. \ref{Fig:FEROS}) presents evident variations. The line profiles shown in Fig. \ref{Fig:FEROS} hint to a stable H$\alpha$ emission with two outbursts detected on Feb. 13, 2011, and Apr. 23, 2014. The FEROS spectrum obtained on Feb. 13, 2011, shows some line variations, but these are most likely due to problems in the continuum normalisation. Despite this last issue, we retained the Feb. 13, 2011, spectrum in this study because of its relevance for the H$\alpha$ line profile variations.

\begin{table}
\centering
\caption{HD\,54879 radial velocities between 2009 and 2014 obtained with three different spectrographs (see Sect. \ref{var}).}
\begin{tabular}{ l r r r r}
\hline
   ID    &            Date &                 HJD- &                  Vrad. \\
           &                 &              2450000 &         (km\,s$^{-1}$) \\
\hline\hline\\
     HARPS &      23-Apr-2014 &   6770.4652        &         29.5$\pm$ 1.0 \\[2pt]
     FIES &      16-Feb-2013 &    6339.5189       &          29.0$\pm$ 3.0 \\[2pt]
     FIES &      26-Dec-2012 &    6287.6420         &        29.0$\pm$ 1.0 \\[2pt]
     FEROS &      19-May-2012 &  6067.4686         &         29.0$\pm$ 2.0 \\[2pt]
     FEROS &      17-May-2011 &  5699.4522        &          29.0$\pm$ 2.0 \\[2pt]
     FEROS &      22-Mar-2011 &  5642.5125        &          30.0$\pm$ 2.0 \\[2pt]
     FEROS &      13-Feb-2011 &  5605.6578        &          30.5$\pm$ 3.0 \\[2pt]
     FIES &      15-Jan-2011 &   5576.5463        &          28.5$\pm$ 2.0 \\[2pt]
         FEROS &      01-May-2009 &  4953.4889        &          29.0$\pm$ 2.0 \\[2pt]
\hline
\hline
\end{tabular}
\label{vra}
\end{table}

 
\subsection{Magnetic field--peculiarities  links}
\label{evo}

\begin{figure}
                \resizebox{\hsize}{!}{\includegraphics[angle=0,width=\textwidth]{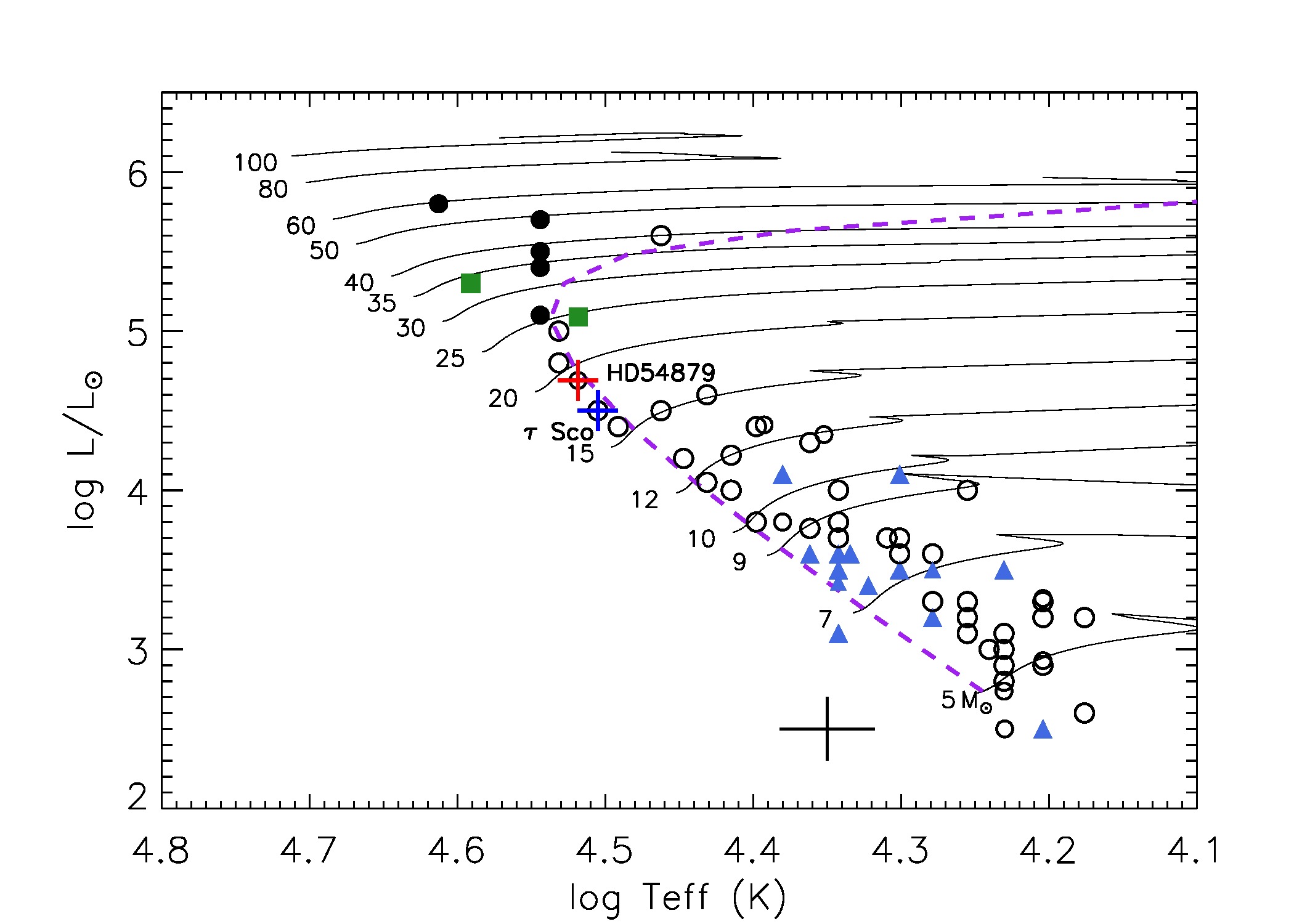}}   
                
                \caption{Position in the Hertzsprung--Russell diagram of OB-stars with a confirmed magnetic field detection  \citep{2013A&A...557L..16B,2013MNRAS.429..398P,2014arXiv1404.5508A,2014A&A...562A.143F,Fossati2014,2014A&A...563L...7N} and \cite{2011A&A...530A.115B} evolutionary tracks.  Stars with an Of?p spectral classification are highlighted with black dots. Other peculiar O-stars are indicated by green squares. HD\,54879 is labelled and shown with an additional red cross. Blue triangles indicate B stars classified as peculiar in the works cited. An isochrone close to the age of HD\,54879, $4.0\,$Myr, is plotted (purple dashed line). The averaged uncertainties  in effective temperature and luminosity are illustrated by the black cross in the bottom part of the plot. The magnetic star in the Trifid nebula detected by \cite{2014A&A...564L..10H} is not included here because of the difficulties in establishing its properties. $\tau$\,Sco has also been labelled and marked (blue cross).}                    
   \label{Fig:sHRD}
        \end{figure}

Figure~\ref{Fig:sHRD} shows the position of the known magnetic chemically peculiar B-type stars (Bp stars) in the HRD. Chemical peculiarities in these objects are closely linked to the magnetic field \cite[e.g.][]{Berger1956,1977A&AS...30...11P,1979ApJ...228..809B,2011ApJ...726...24K,2014A&A...561A.147B} and arise as a result of diffusion processes (i.e. that depend on the balance between gravitational settling and radiative levitation; \citealt{1970ApJ...160..641M}). The magnetic B-type stars falling in the same region of the HRD as the Bp stars are also expected to present chemical peculiarities \cite[see e.g.][]{2014arXiv1404.5508A}, but a detailed abundance analysis has not been performed yet for all of them. The stellar wind sets the upper limit in temperature and mass at which a B-type star may develop surface chemical peculiarities as a result of diffusion and, in theory, a plot such as Fig.~\ref{Fig:sHRD} would allow  these important boundaries to be determined. In practice, this is complicated by the fact that the wind is line-driven and therefore the surface abundances \citep{2014A&A...564A..70K}, as well as the magnetic field, control the wind strength. As stars may have different magnetic field (strengths and topologies) and surface abundances (still to be determined in several cases), it is not yet possible to firmly constrain the upper temperature and mass boundary of diffusion in magnetic stars. Still, Fig. \ref{Fig:sHRD} suggests a boundary at about 10\,M$_\odot$ and 25000\,K.

 HD\,54879 seems to show remarkable differences compared to the other known magnetic O-type stars. The star does not present the spectral features typical of Of?p stars and, in addition,  it does not present the peculiarity and variability displayed by HD\,37022 ($\theta^1$\,Ori\,C) \citep{1993A&A...274L..29S,1994ApJ...425L..29W}. The other magnetic O-type stars also present peculiarities of some sort and spectral variability: Plaskett's star appears to be the mass gainer in the HD\,47129 binary system \citep[][]{2013MNRAS.428.1686G}, and HD\,37742 ($\zeta$\,Ori\,Aa) is an evolved rapidly rotating O-type star with a very weak magnetic field \citep[anomalous for magnetic massive stars;][]{2008MNRAS.389...75B}. A wide variety of spectral variability for the apparently normal stars ALS\,15128 \citep{2012MNRAS.423.3413N,2011BSRSL..80..644C} and HD\,57682 \citep{2009MNRAS.400L..94G,2012MNRAS.426.2208G} has also been reported. HD\,54879 instead does not show any spectral peculiarity or line profile variability in the photospheric lines and appears to be the only known magnetic single O dwarf to date with an apparently normal and stable photospheric spectrum.  \cite{2010ApJ...711L.143W} \citep[see also][]{1972AJ.....77..312W} classified the Of?p stars mainly according to the presence of C\,{\sc iii} $\lambda\lambda4647-4650-4652\,\AA$ emission lines. It could be that HD\,54879 is simply not hot and luminous enough to display the morphological features of the stars belonging to the Of?p class.

\subsection{Similarities with $\tau$\,Sco}
\label{tauScor}
 The optical spectrum of HD\,54879 resembles that of $\tau$\,Sco. Both stars have a low \ensuremath{{\upsilon}\sin i} and a  similar effective temperature and surface gravity \citep{2012A&A...539A.143N},  see Fig. \ref{Fig:sHRD}. However, HD\,54879 does not share other peculiarities reported for $\tau$\,Sco, particularly the relatively high nitrogen-to-carbon ratio presented by $\tau$\,Sco  is not found in HD\,54879. \cite{2014A&A...566A...7N} claim that $\tau$\,Sco is a blue straggler that has been rejuvenated either by a merger or by mass accretion in a binary system. The agreement between the age derived from {\sc bonnsai} and that of the CMa OB1 association suggests that HD\,54879 is not a blue straggler (i.e. it has not been rejuvenated). There might also be a difference in the magnetic field topology, but more polarimetric observations of HD\,54879 are required to conclude anything on this. 

\cite{2011MNRAS.412L..45P} presented two stars that appear to be analogues to  $\tau$\,Sco based on their $T_{\rm{eff}}$ and $\rm{log}\,$g values, the detection of magnetic fields, and  UV spectral characteristics. There are no UV data available for HD\,54879. 

\section{Conclusions}           

We report  the first measurement of a magnetic field in HD\,54879, a presumably single O9.7~V star with a low projected rotational velocity, observed in the context of the BOB collaboration. Based on HARPS and FORS\,2 spectropolarimetric data we characterised the stellar atmosphere of HD\,54879 and unambiguously detected a strong magnetic field using independent techniques.  The magnetic field measurements were carried out in two independent ways, and  reached consistent values. We derived a lower limit on the polar magnetic field of $\sim$\,$2.0\,$kG.
 
We analysed the optical spectra of HD\,54879 using  {\sc fastwind} grids and automatic routines. Both FORS\,2 and HARPS datasets yield almost identical effective temperature and surface gravity. The same is true for the chemical abundances. The chemical composition is systematically  lower than  the solar one \citep{2009ARA&A..47..481A} and lower than the cosmic abundance standard from \cite{2012A&A...539A.143N}, but compatible with both within the uncertainties; neither obvious enhancements nor depletions were found.  The match between low  and high spectral resolution analyses supports the robustness of our results and lends confidence to previous quantitative analyses based on low spectral resolution data.

The mass-loss rate derived based on the H$\alpha$ emission is unexpectedly  large for an O9.7~V star. A comparison with a standard star of a similar spectral type (HD\,36512), and with previous works on O9-B0 dwarfs confirmed this, rejecting  H$\alpha$ as a reliable mass-loss rate indicator for HD\,54879. Our analysis suggests that  circumstellar material is a more plausible explanation for the H$\alpha$ emission. The measurable differences in the Balmer lines and He\,{\sc ii} $\lambda4686\,\AA$ are also ascribed to circumstellar material.

We explored line profile variability using high spectral resolution data from the IACOB and OWN surveys. We checked for changes  in nine spectra covering six years. The optical photospheric transitions remain  unchanged. The  H$\alpha$  emission displays a fairly stable shape with an enhanced emission in the line core in 2011 and 2014. This could be an indication for periodical outburst events, but more data are needed to establish this.

The optical spectrum resembles that of $\tau$\,Sco, but unlike $\tau$\,Sco itself, HD\,54879  presents surface abundances compatible with the solar and the standard cosmic abundances. In addition, the age of HD\,54879  matches  the age of the CMa OB1 association, to which this star probably belongs. These considerations make the blue straggler hypothesis unlikely and suggest that the magnetic field was not generated by a merger  of two main-sequence stars.

 HD\,54879 is, so far, the strongest magnetic single O-type star detected with a stable and normal optical spectrum, with the exception of the lines partly formed in the magnetosphere.  We have not detected any distinctive spectral feature observed in other magnetic O-type stars (i.e. Of?p objects). Nonetheless, it may be a consequence of the star's lower temperature and luminosity compared to the known Of?p stars.

HD\,54879 is certainly an interesting object to follow up. The strong magnetic field makes this star a good candidate for exploring the apparent ordered magnetic geometry and H$\alpha$  variability. The apparent  link between  H$\alpha$ emission, not expected according to its spectral type, and a magnetosphere offers a  criterion for selecting magnetic candidates.

\begin{acknowledgements}
The authors thank the referee for useful comments and helpful suggestions that improved this manuscript. LF acknowledges financial support from the Alexander von Humboldt Foundation.
SS-D and AH thank funding from the Spanish Government Ministerio de Economia y Competitividad (MINECO) through grants AYA2010-21697-C05-04, AYA2012-39364-C02-01 and Severo Ochoa SEV-2011-0187, and the Canary Islands Government under grant PID2010119. TM acknowledges financial support from Belspo for contract PRODEX GAIA-DPAC. FRNS acknowledges the fellowship awarded by the Bonn--Cologne Graduate School of Physics and Astronomy. NL and AR thank the DFG (Germany) and CONICYT (Chile) for the International Collaboration Grant DFG-06. The authors thank Andreas Irrgang for helping in the FEROS  data reduction.

\end{acknowledgements}



\bibliographystyle{aa}

\bibliography{hd54879}
\Online 
\begin{appendix}
\section{HARPS best-fit modelling}
\begin{figure*}
                \resizebox{\hsize}{!}{\includegraphics[angle=0,width=\textwidth]{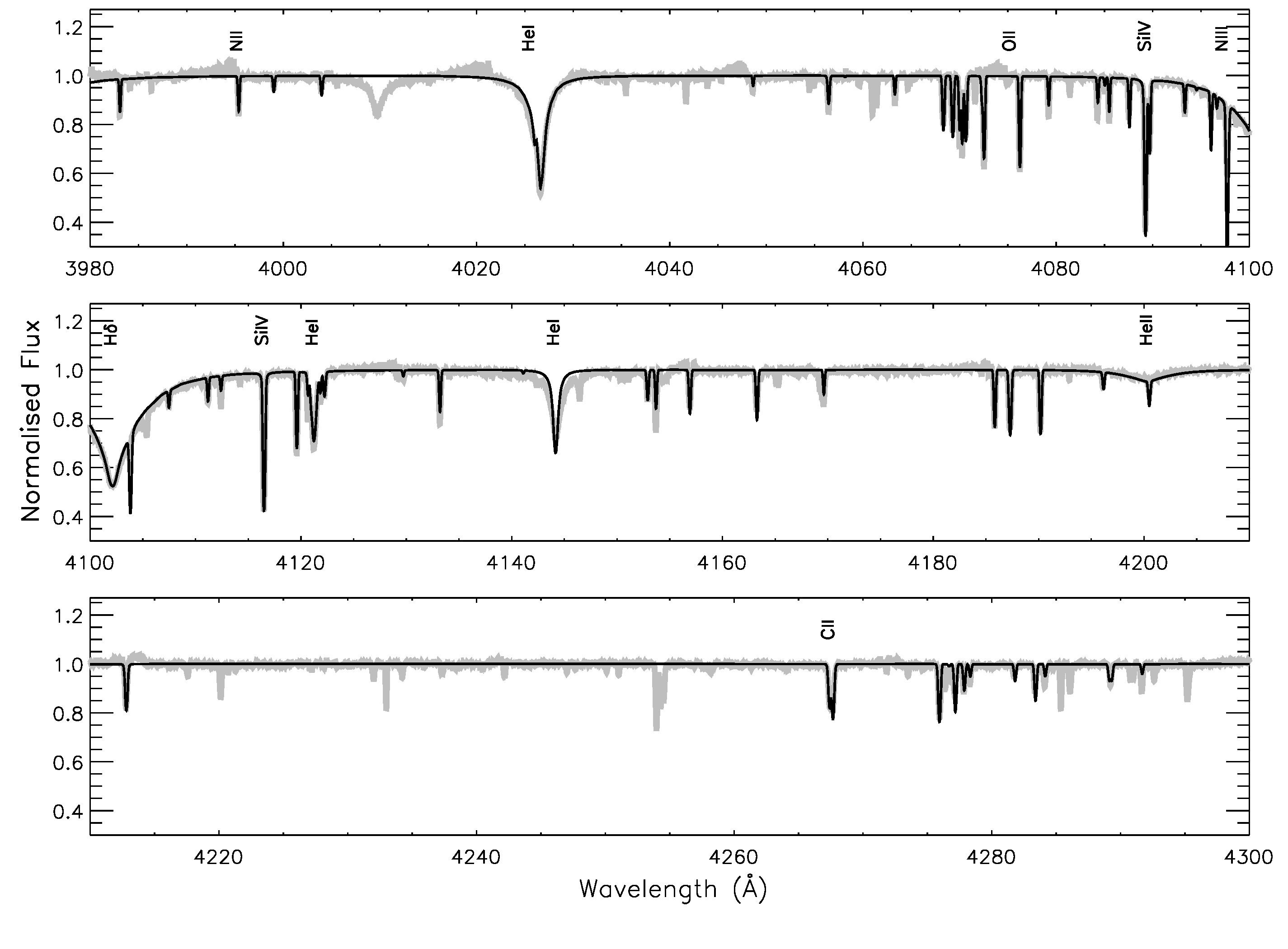}}
                \caption{Normalised optical spectrum (grey) and the best-fit \textsc{fastwind} stellar model (black), see Fig.~\ref{Fig:HARPS}.}
                \label{Fig:HARPSA1}
        \end{figure*}
        
  \begin{figure*}
                \resizebox{\hsize}{!}{\includegraphics[angle=0,width=\textwidth]{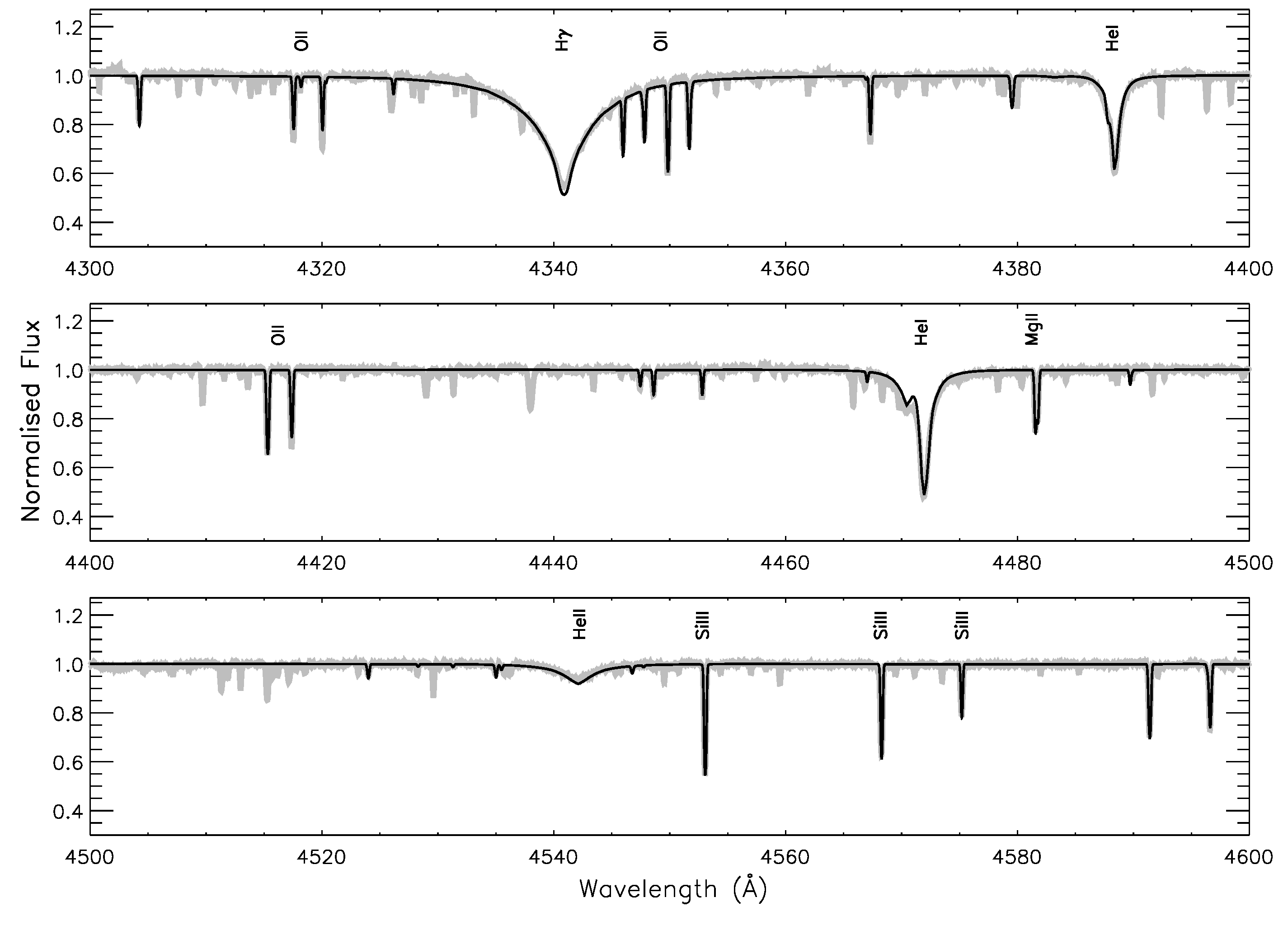}}
                \caption{Figure \ref{Fig:HARPSA1} continued.}
                \label{Fig:HARPSA2}
        \end{figure*}
        
    \begin{figure*}
                \resizebox{\hsize}{!}{\includegraphics[angle=0,width=\textwidth]{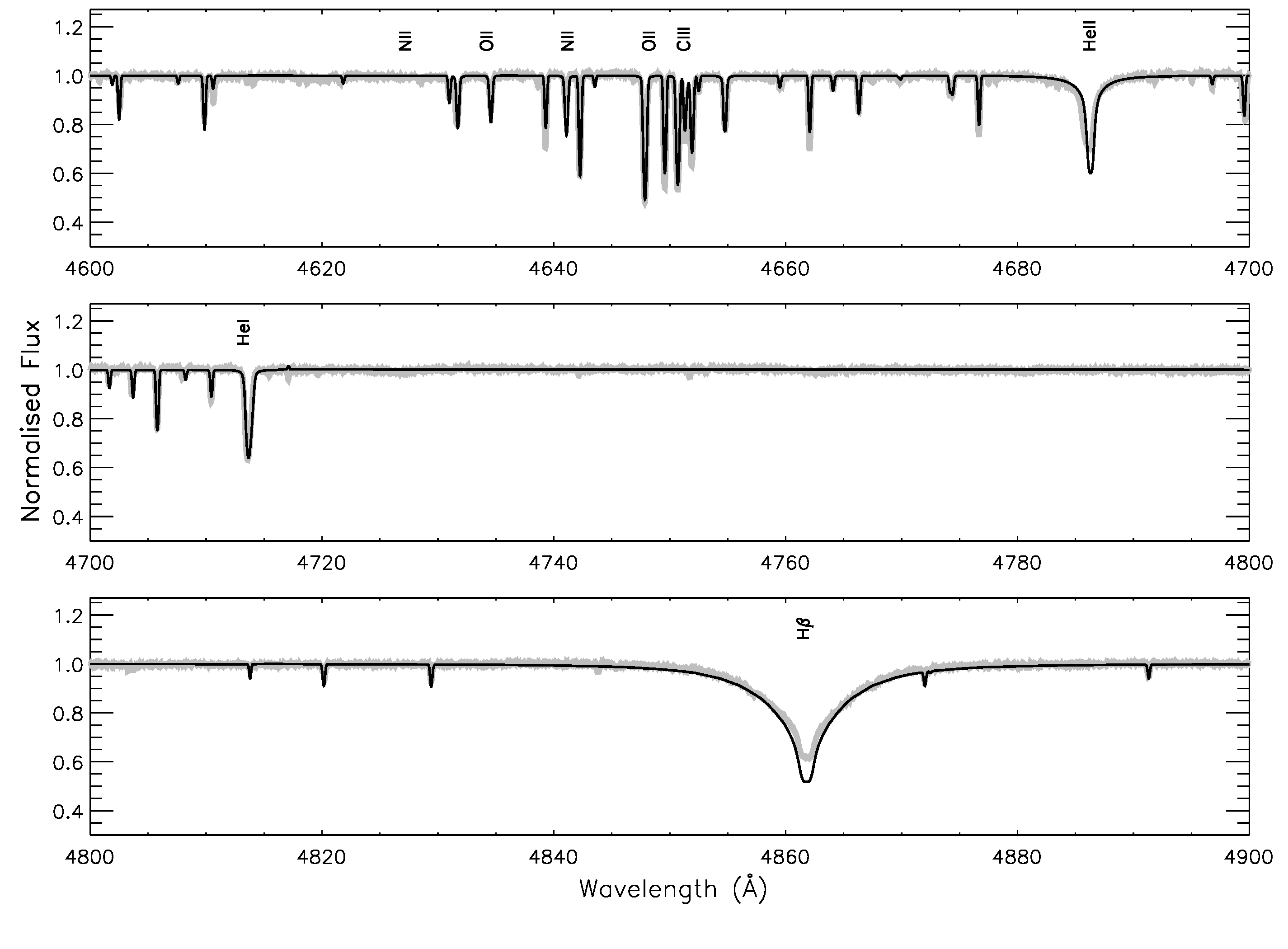}}
                \caption{Figure \ref{Fig:HARPSA1} continued.}
                \label{Fig:HARPSA3}
        \end{figure*}
        
  \end{appendix}

\end{document}